\documentclass[aps,pre,reprint]{revtex4-1}
\usepackage{amsmath, amssymb}
\usepackage[latin9]{inputenc}
\usepackage{graphicx,epstopdf}
\usepackage{enumerate}
\usepackage{multirow}
\usepackage{subfigure}
\usepackage{dsfont}
\usepackage{slashed}
\usepackage{textcomp}
\usepackage{url}
\usepackage{braket}

\begin{document}
\title{Free Energy Barriers for Crystal Nucleation from Fluid Phases}
\author{Peter Koß$^{1,2}$}
\author{Antonia Statt$^{1,2,3}$}
\author{Peter Virnau$^{1}$}
\author{Kurt Binder$^{1}$}
\affiliation{$^1$Institut für Physik, Johannes Gutenberg-Universität, D-55128 Mainz, Staudinger Weg 9, Germany}
\affiliation{$^2$Graduate School Materials Science in Mainz, D-55128 Mainz, Staudinger Weg 9, Germany}
\affiliation{$^3$Department of Chemical and Biological Engineering, Princeton School of Engineering and Applied Science, Princeton, NJ 08544}
\date{\today}
\begin{abstract}
Monte Carlo simulations of crystal nuclei coexisting with the fluid phase in thermal equilibrium in finite volumes are presented and analyzed, for fluid densities from dense melts to the vapor. Generalizing the lever-rule for two-phase coexistence in the canonical ensemble to finite volume, ``measurements'' of the nucleus volume together with the pressure and chemical potential of the surrounding fluid allows to extract the surface free energy of the nucleus. Neither the knowledge of the (in general non-spherical) nucleus shape nor of the angle-dependent interface tension is required for this task. The feasibility of the approach is demonstrated for a variant of the Asakura-Oosawa model for colloid-polymer mixtures, which form face-centered cubic colloidal crystals. For a polymer to colloid size ratio of $0.15$, the colloid packing fraction in the fluid phase can be varied from melt values to zero by the variation of an effective attractive potential between the colloids. It is found that the approximation of spherical crystal nuclei often underestimates actual nucleation barriers significantly. Nucleation barriers are found to scale as $\Delta F^*=(4\pi/3)^{1/3}\bar{\gamma}(V^*)^{2/3}+const.$ with the nucleus volume $V^*$, and the effective surface tension $\bar{\gamma}$ that accounts implicitly for the nonspherical shape can be precisely estimated.
\end{abstract}
\maketitle

\section{Introduction and Overview\label{sec:intro}}
\begin{figure}
\centering
\includegraphics[width=0.5\textwidth]{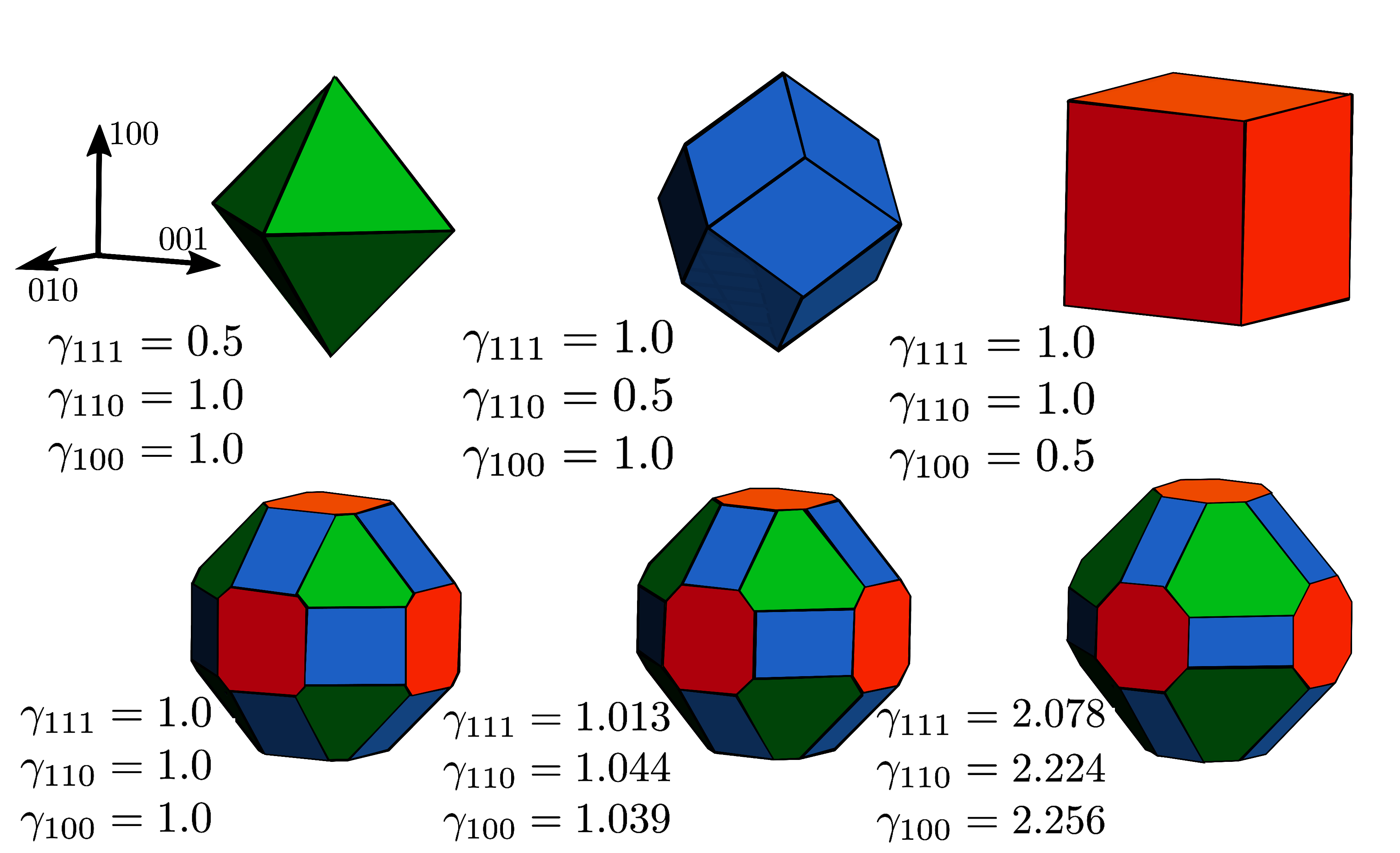}
\caption{\label{fig:wulff}Three dimensional restricted Wulff shapes of a fcc nucleus assuming only values for the three interfacial tensions $\gamma_{111}$, $\gamma_{100}$ and $\gamma_{110}$, allowing only the (111), (100) and (110) orientations of the vector perpendicular to the interface. Arbitrarily chosen values of $\gamma_{111}$, $\gamma_{100}$ and $\gamma_{110}$ are indicated together with the resulting crystal shapes. The top row shows cases where only a single type of interface results (octahedron, rhombic dodecahedron, and cube); (111) interfaces are shown in green, (110) interfaces in blue, and (100) interfaces in red. The lower row shows shapes with all three faces. Note that lower middle and right figure uses choices for the interfacial tensions as they did result for our numerical model study. These shapes were all constructed using a software published by Zucker et al. \cite{14}.}
\end{figure}

\begin{figure*}
\centering
\includegraphics[width=0.9\textwidth]{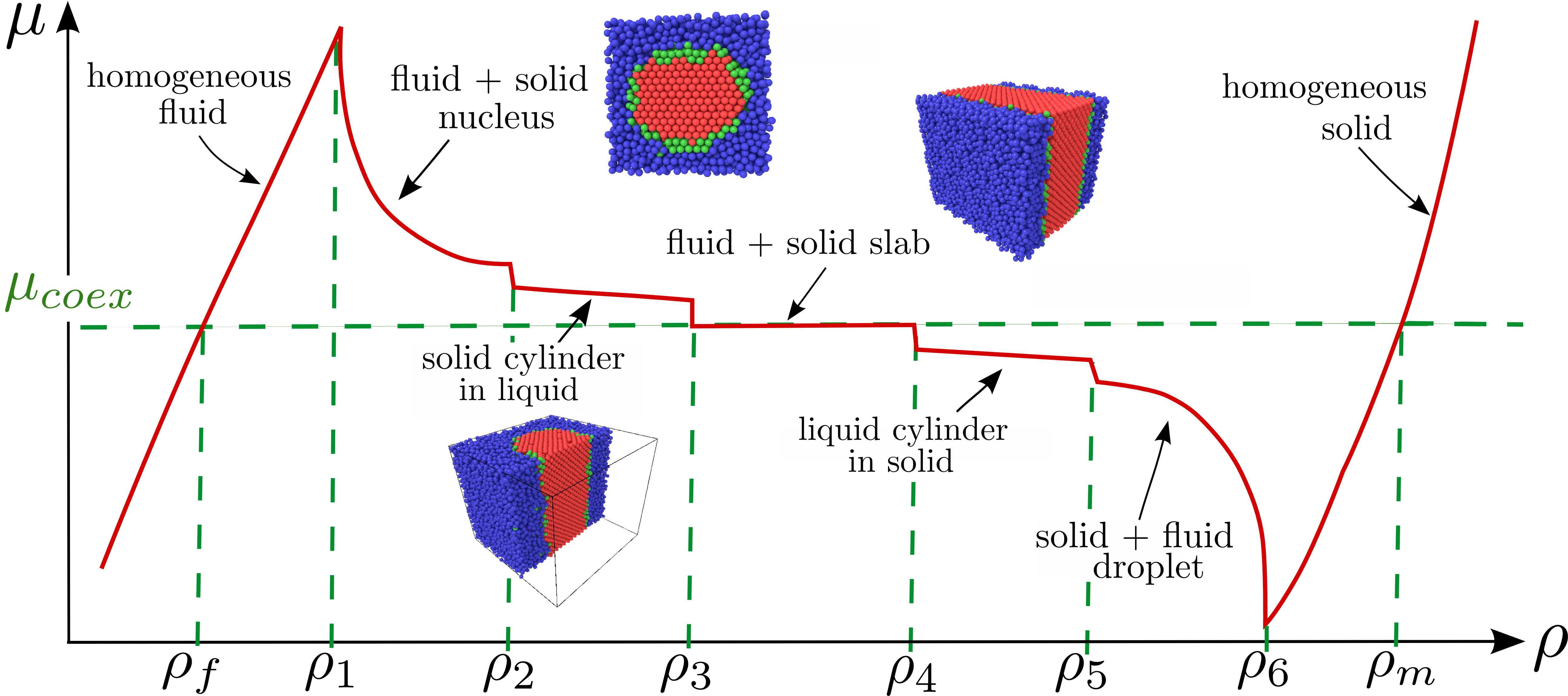}
\caption{\label{fig:chvsrho}Schematic plot of the chemical potential $\mu$ vs. density $\rho$ for a system undergoing a phase transition from a liquid to a crystal, comparing the behavior in the thermodynamic limit to the case of finite size of the volume $V$, assuming PBC. For $V\rightarrow\infty$, $\mu$ increases with $\rho$ until the freezing density $\rho_f$ is reached, then $\mu=\mu_{coex}=const.$ up to $\rho=\rho_m$, and then $\mu$ increases further in the homogeneous crystal. For finite $V$ the homogeneous fluid is stable up to the density $\rho_1$ where the droplet evaporation/condensation transition occurs. The descending parts of the $\mu(\rho)$ vs. $\rho$ curve reflect interfacial effects which are negligible in the limit $V\rightarrow\infty$ only. Coexistence between compact nuclei und surrounding fluid occurs for $\rho_1<\rho<\rho_2$, while for $\rho_2<\rho<\rho_3$ the nucleus is cylindrical (stabilized by the PBC), and for $\rho>\rho_3$ slab-like domain configurations take over. The color code of the corresponding simulation snapshots indicates particles in the fluid as blue and particles in the crystal as red, particles in the topmost surface layer of the crystal are colored green.}
\end{figure*}

Understanding nucleation phenomena in condensed matter is a fundamental problem of statistical mechanics and has important applications, for atmospheric sciences (formation of raindrops or snow crystals), geosciences (crystallization of silicate melts), and material science (e.g. microstructure of metallic alloys obtained by cooling their melts), for instance \cite{1,2,3,4,5,6}. The classical theory assumes that nanoscopic nuclei of the crystalline solid form spontaneously by thermal fluctuations when one suddenly has brought the system from the region where the fluid phase is the thermodynamically stable phase beyond the fluid-solid coexistence curve, so that the fluid phase is only metastable (``homogeneous nucleation''). Although in the metastable region the bulk free energy density of the crystal is lower than the free energy density of the metastable fluid, large-scale phase transformation from fluid to solid is often prevented by the cost in surface excess free energy caused by the fluid-solid interface. This means a nucleation event amounts to overcoming a free energy barrier. The classical theory of homogeneous nucleation estimates this barrier assuming that the formation free energy of the nucleus can be written in terms of macroscopic nucleus volume ($V_n$) and surface terms,
\begin{equation}
\Delta F^*(V_n)=-(p_c-p_l)V_n+F_{surf}(V_n)~.
\end{equation}
Here $p_c$ is the pressure inside of the crystal nucleus, and $p_l$ the pressure in the metastable liquid phase surrounding it. The traditional basic assumption of the theory then neglects the fact that for a crystal-fluid interface the interfacial tension $\gamma(\vec{n})$ depends on the orientation of the interface normal ($\vec{n}$) relative to the crystal axes. If one could replace $\gamma(\vec{n})$ by a constant $\bar{\gamma}$, assuming that the nucleus has a spherical shape, we could write ($R_n$ is the radius of the nucleus)
\begin{equation}
F_{surf}(V_n)=4\pi R_n^2\bar{\gamma}=A_{iso}\bar{\gamma}V_n^{2/3}~,~A_{iso}=(36\pi)^{1/3}~.
\end{equation}
When the anisotropy of the interfacial tension is taken into account, however, for large $V_n$ the equilibrium shape of a crystal nucleus clearly is not a sphere, but rather given by the Wulff construction \cite{7,8,9,10,11}. In fact, at low temperatures (T lower than the appropriate roughening transition temperatures $T_R$ \cite{12,13}) crystal shapes with strictly planar facets occur. As an extreme example, Fig. \ref{fig:wulff} shows crystal shapes of face-centered cubic crystals that result \cite{14} when only three interface orientations (111), (100), and (110) would be permitted (recall that an (111) interface in the ideal lattice structure is a closed packed plane with triangular lattice structure, and it is natural to expect that the interface tension $\gamma_{111}$ of this surface is smaller than $\gamma_{100}$ or $\gamma_{110}$).\\
While shapes as shown in Fig. \ref{fig:wulff} are expected as $T\rightarrow 0$, in general one expects to have facets with rounded corners and edges (for nonzero $T<T_R$) and nonspherical but curved shapes for $T>T_R$. An example for the gradual change of a crystal shape with increasing temperature has been presented for the nearest neighbor simple cubic Ising model recently \cite{15}: there occurs a gradual change from a cube (such as shown in the rightmost figure in the top row of Fig. \ref{fig:wulff}) at $T=0$ to a sphere which only occurs for temperatures very close to the critical point of this lattice model. Of course, this model completely lacks an off-lattice fluid phase, and hence this model cannot capture the essential features of the crystal/fluid transition.\\
When the interface tension is anisotropic, one can in principle generalize Eq. (2) by writing the surface free energy $F_{surf}(V_n)$ as an integral over the crystal surface,
\begin{equation}
F_{surf}(V_n)=V_n^{2/3}\int_{A_W}\gamma(\vec{n})d\vec{s}\equiv A_W\bar{\gamma}V_n^{2/3}~,
\end{equation}
where $A_W$ is the surface area of a crystal having the shape resulting from the unrestricted Wulff construction \cite{7,8,9,10,11} and unit volume. Note that the geometric factor $A_W$ must always exceed $A_{iso}$ (e.g., for the cube resulting in the Ising model $A_W=6$, $A_W/A_{iso}=(6/\pi)^{1/3}$), and the average interfacial tension
\begin{equation}
\bar{\gamma}=(A_W)^{-1}\int_{A_W}\gamma(\vec{n})d\vec{s}
\end{equation}
in general will exceed the smallest of the interfacial tensions, when more than one type of interface contributes.\\
Now, according to the classical theory of homogeneous nucleation, one simply combines Eqs. (1) and (2) or (3) to identify the nucleation barrier from the maximum of $\Delta F(V_n)$. This yields the nucleus linear dimension at the barrier
\begin{equation}
V_n^{*^{1/3}}=\frac{2A_W\bar{\gamma}}{3(p_c-p_l)}
\end{equation}
and the associated free energy barrier
\begin{equation}
\Delta F^*=\frac{1}{3}A_W\bar{\gamma}V_n^{*^{2/3}}=\frac{1}{2}(p_c-p_l)V_n^*~.
\end{equation}
The nucleation rate then is assumed to be proportional to 
\begin{equation}
J\propto exp(-\Delta F^*/k_BT)~,
\end{equation}
where nonexponential prefactors occur describing both kinetic effects and corrections due to various fluctuation effects. The nucleation rate is defined as the number of critical nuclei (with volume $V_n^*$) formed per unit volume and unit time in the limting case where the metastable state has had enough time to reach at least local equilibrium, and nucleation events are seldom enough so the change of this state due to the forming nuclei can be neglected, and nucleation can be approximated as a steady-state process \cite{1,2,3,4,5,16,17,18,19}.\\
Fluctuation effects are important, since for the case of barriers that are of physical interest ($\Delta F^*<10^2k_BT$) critical nuclei are nanoscopically small objects. Eqs.(1-6) describe the nuclei in terms of a simple variable $V_n$ and then $\Delta F(V_n)$ has just a simple maximum at $V_n^*$, however, in reality nuclei have many degrees of freedom, and instead of a maximum of a simple function one deals with a saddle point of the coarse-grained free energy density in a high-dimensional configuration space \cite{18,20,21}. Some of the resulting fluctuation corrections are commonly attributed to capillary wave excitations of the interface \cite{22}. The latter occur only for rough fluid-crystal interfaces, of course, but have been observed in computer simulations of various models \cite{23,24,25,26,27} and their analysis is useful for the estimation of the interfacial stiffness $\tilde{\gamma}$ of these interfaces. While such fluctuation corrections are typically of order $ln(V_n)$, a curvature correction of order $V_n^{1/3}$ is also controversially discussed (mostly for the case of vapor-liquid interfaces, e.g. \cite{28,29,30,31,32,33}), sign and magnitude of this correction are uncertain.\\
While for nucleation of liquid droplets from the vapor the two phases differ by the value of a scalar order parameter, the density, it is rather natural to use a description of the nucleus in terms of a simple variable as a starting point, namely the nucleus radius or volume. However, for the liquid to solid transition the distinction of the phases in terms of order parameters is very different: with crystallization both the translational and orientational symmetry of the liquid is broken \cite{34}, and there is an accompanying density change (see e.g. \cite{35} for a simulation study of synchronized ordering and densification).
Clearly, such considerations must enter a first-principles discussion of crystal nucleation, but this is beyond the scope of the present work, which focusses on the effect that the nucleus shape is in general nonspherical.\\
As a consequence, quantitatively reliable predictions of nucleation barriers are difficult. For the vapor-liquid transition the nucleus on average has a spherical shape, and the interface tension $\gamma$ can be reliably measured (in the case of experiments) or predicted from models for the intermolecular forces (in the case of computer simulations), and thus the limiting case of the classical theory for the nucleation barrier (Eq. (6) with $A_W=A_{iso}$, cf. Eq. (2)) of large enough nuclei can be easily worked out. For the liquid-solid transition, however, $\gamma(\vec{n})$ and hence $\tilde{\gamma}$ cannot be easily measured in experiments, and finding it in simulations is a major effort and only possible approximately (e.g. \cite{27,36a}). Thus, working out \linebreak $\Delta F^*\{$Eq.$\,$(6)$\}$ according to the classical nucleation theory is problematic, and hence tests of the theory by experiment can hardly yield conclusive results (see e.g. \cite{36}).\\
In this situation, a computer simulation approach to the problem appears as the most promising choice, but it also encounters conceptual problems: many techniques for obtaining $\Delta F^*$ of a nucleus require a distinction which particles are part of the crystal and which particles are part of the surrounding fluid \cite{37,38}. There is some ambiguity concerning the particles in the interfacial region at the surface of the nucleus, however. Moreover, a single critical nucleus surrounded by  ``supersaturated''  (or supercooled, respectively) fluid at constant pressure in a macroscopically large volume is intrinsically unstable. As an object just at the top of the nucleation barrier it is bound to either grow or decay: thus performing a sampling of its properties one usually needs to apply a biasing potential \cite{30,37} to stabilize the nucleus at a pre-chosen size.\\
In the present paper, we present an alternative simulation approach, where the stable equilibrium between crystal nucleus and surrounding liquid in a finite volume is studied in the NVT ensemble. This stable equilibrium is a consequence of the lever rule \cite{39} which describes two-phase coexistence in the canonical ensemble in the thermodynamic limit: when the density $\rho$ of the particles in the volume V exceeds the density $\rho_f$ describing the onset of freezing, but is still much smaller than the density $\rho_m$ describing the onset of melting of a crystal in equilibrium, we must have a crystal of volume $V_n$ such that ($N=\rho V$ while $N_n=\rho_mV_n$)
\begin{equation}
\rho V=\rho_mV_n+\rho_f(V-V_n)~.
\end{equation}
Eq. (8) holds in the limit $V\rightarrow\infty$, $N\rightarrow\infty$ taken such that $\rho_f<\rho<\rho_m$, and both pressure $p$ and chemical potential $\mu$ are constants independent of $\rho$,
\begin{equation}
p(\rho,T)=p_{coex}=p_l(\rho_f,T)=p_c(\rho_m,T)~,
\end{equation}
\begin{equation}
\mu(\rho,T)=\mu_{coex}=\mu_l(\rho_f,T)=\mu_c(\rho_m,T)~.
\end{equation}
Here we have denoted the branches of the pressure $p(\rho,T)$ in the pure phases as $p_l(\rho,T)$ in the liquid and $p_c(\rho,T)$ in the crystal, and in the same way for $\mu$ ($\mu_l(\rho,T)$, $\mu_c(\rho,T)$). When one generalizes Eq. (8) to a finite volume, we need to specify the boundary conditions (we shall consider periodic boundary conditions (PBC) throughout) and the convention used for dividing up the particles between crystal nucleus and surrounding fluid: we use here the Gibbs convention of an equimolar dividing surface, so that $N=N_f+N_n$, $N_f$ being the particle number in the fluid and $N_n$ the particle number in the nucleus, no particle number excess associated with the interface being permitted. Then Eq. (8) still holds, but the applicable density range is restricted to $\rho_1<\rho<\rho_2$, where $\rho_1$ is the density where the  ``droplet evaporation/condensation transition'' \cite{40,41,42,43,44,45,46} occurs, while $\rho_2$ is the density where a transition from a compact droplet shape to a cylindrical droplet shape (stabilized by the PBC) occurs (Fig. \ref{fig:chvsrho}).
Now due to nonnegligible interfacial contributions to the thermodynamic potential Eqs. (9),(10) no longer hold, both chemical potential and pressure are enhanced in comparison with their coexistence values,
\begin{align}
\mu(\rho,T)=\mu_l(\rho_l,T)=\mu_c(\rho_c,T)>\mu_{coex}~,\\
p_c(\rho_c,T)>p_l(\rho_l,T)>p_{coex}~.
\end{align}
While in the presence of density inhomogenities in a system and corresponding interfacial contributions to its total free energy the chemical potential in equilibrium must still be spatially homogeneous, the pressure is not (Laplace pressure effect).\\
Due to the shift of equilibrium conditions from $\mu_{coex}$ to the chemical potential $\mu$ as specified in Eq. (11), the coexisting domains in the finite system no longer have the densities $\rho_f$ and $\rho_m$, but also have enhanced values $\rho_l$ and $\rho_c$, as anticipated in Eq. (11),(12), and hence Eq. (8) in the finite system gets replaced by 
\begin{equation}
\rho V =\rho_cV_n+\rho_l(V-V_n)~.
\end{equation}
The idea of the simulation approach described in this paper hence is a numerical analysis of the equilibrium described by Eqs. (11)-(13): choosing $\rho$ and $V$ suitably, a crystal with volume $V_n$ is equilibrated. The densitiy $\rho_l$ of the fluid in its surroundings as well as the associated pressure $p_l(\rho_l,T)$ and chemical potential $\mu_l(\rho_l,T)$ can then be calculated from the simulation. Since $\mu_c(\rho_c,T)=\mu_l(\rho_l,T)$, the density $\rho_c$ can also be extracted (from a suitable thermodynamic integration procedure). Then Eq. (13) yields the associated value of $V_n$ readily. If $V_n$ is large enough such that corrections to Eqs. (1),(2) are negligible, one can readily apply Eq. (6) to obtain the nucleation barrier from $\Delta F^*=\frac{1}{2}(p_c-p_l)V_n^*$, and no consideration of $\gamma(\vec{n})$ and the resulting crystal shape is required. There will be two important consistency checks for the validity of this approach: (i) for large $V_n$ we must have $\Delta F^*\propto V_n^{2/3}$, and (ii) the results for $\Delta F^*$ must be independent from the total volume $V$. Both conditions are not expected to be true if $V$ is too small, when finite size effects related to fluctuations (which have nonzero correlation lengths $\xi_l$, $\xi_c$ in the pure fluid and crystal phases) come into play. Obviously this approach is not well suited, and not intended for, the study of ultrasmall nuclei containing a few particles only; we rather address the prediction that the classical theory of nucleation makes for crystal nucleation barriers. Although for very small nuclei corrections to this classical prediction are to be expected, it nevertheless is a very useful reference result, also for extensions to heterogeneous nucleation (e.g. \cite{47,48,49,50,51,52,53,54,55,56}). The classical approach there considers sphere-cap shaped droplets on planar substrates under conditions of partial wetting \cite{57a}, and clearly for nucleation of crystals interfacial tension anisotropies are again expected to matter \cite{55}.\\
In section \ref{sec:softEffAO} of the present paper we shall describe the model which we use to test the practical applicability of the approach outlined above, namely a version of the well-known Asakura-Oosawa model \cite{57,58,59,60,61} of colloidal particles immersed in a fluid where added polymers provide a tunable effective attractive potential among the colloids \cite{62}. Unlike the original model where the colloid-colloid interaction is a hard-core potential, we use a variant with a strongly repulsive continuous potential \cite{63}, the so-called softEffAO model \cite{63,64}. The advantage of this model is that the pressure in the liquid phase, which is of crucial importance in our analysis, can be straightforwardly computed from the virial formula \cite{65}, which cannot be applied for hardcore potentials. As far as a possible comparison of our results to real colloid-polymer mixtures [66] would be concerned, we note that real colloids anyway never interact strictly with hard-core potentials \cite{67,68}. Using Monte Carlo simulations \cite{69,70,71} in the $NpT$ ensemble with PBC, the equation of state of homogeneous fluid and crystal phases is readily obtained very accurately, varying the temperature-like control parameter over a wide range.\\
Sec. \ref{sec:bulkphase} then discusses the precise estimation of coexistence conditions. We compare results from a method based on kinetics of a ``slab'' configuration (if $p\neq p_{coex}$ the slab grows or shrinks) \cite{25,26} to finite size extrapolations of the equilibrium described by Eqs. (11)-(13) to the thermodynamic limit, and find both approaches in very satisfactory agreement.\\
Sec. \ref{sec:equilibrium} describes the preparation of the systems containing a nucleus surrounded by fluid in thermal equilibrium and discussses their analysis. It is shown that $\Delta F^*(V_n^*)$ indeed is nicely consistent with Eq. (6), and does allow a rather accurate estimation of the effective surface tension $\gamma_{\text{eff}}=(A_W\bar{\gamma})/A_{iso}$ of the fluid-crystal interface. Evidence is presented that $\gamma_{\text{eff}}$ exceeds the surface tension $\gamma_{111}$ of the close-packed (111)-interface of the face-centered cubic crystal distinctly. Finally Sec. \ref{sec:conc} contains some conclusions and gives an outlook on open problems.
\section{The soft effective Asakura-Oosawa (SoftEffAO) model and its bulk phase behavior\label{sec:softEffAO}}
\begin{figure}
\centering
\includegraphics[width=0.45\textwidth]{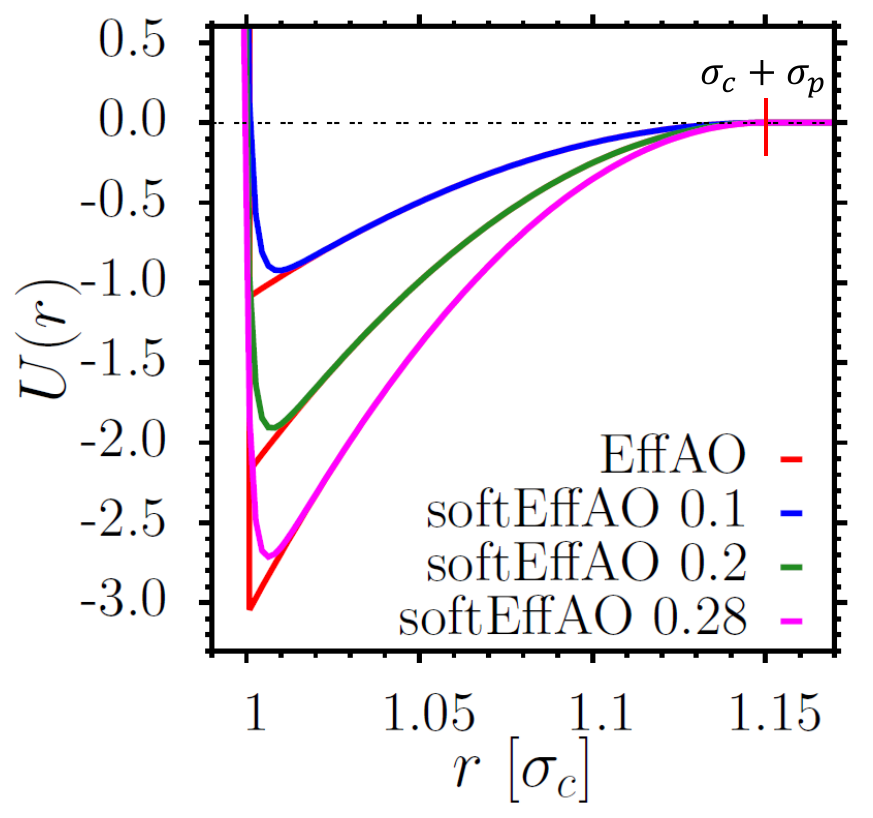}
\caption{\label{fig:softeffAO}Potential $U(r)$ [in units of $k_BT$] vs. distance $r$ [in units of $\sigma_c$] for the softeffAO model, for the case $q=0.15$, and three choices of $\eta_p^r$, $\eta_p^r=0.1$, 0.2, and 0.28. The corresponding potentials $U_{\text{eff}}(r)$ for the original effective AO-model are included for comparison.}
\end{figure}

\begin{figure}
\centering
\includegraphics[width=0.4\textwidth]{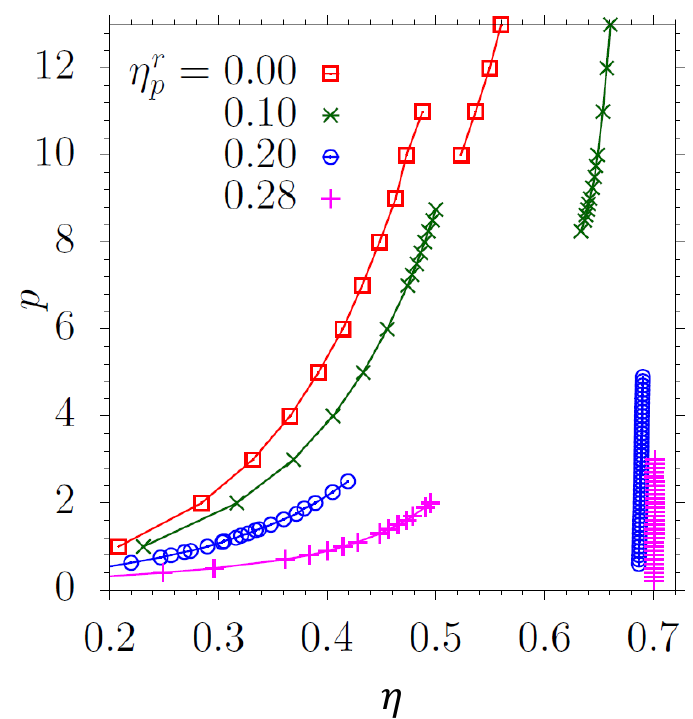}
\caption{\label{fig:eos}Equation of state of the softEffAO model for $q=0.15$ and for four choices of $\eta_p^r$ [Eq. (16)] as indicated. Curves connecting the data points are guides to the eye only, and the statistical errors are by far smaller than the size of the symbols. Data for $\eta<0.5$ denote the fluid branch $p_l(\eta)$ while data for $\eta>0.5$ denote the crystal branch $p_c(\eta)$ of the corresponding isotherms. Pressure is always measured in units of $k_BT/\sigma_c^3$ (with the unit of length $\sigma_c=1$, and $k_BT=1$).}
\end{figure}

\begin{figure}
\centering
\includegraphics[width=0.45\textwidth]{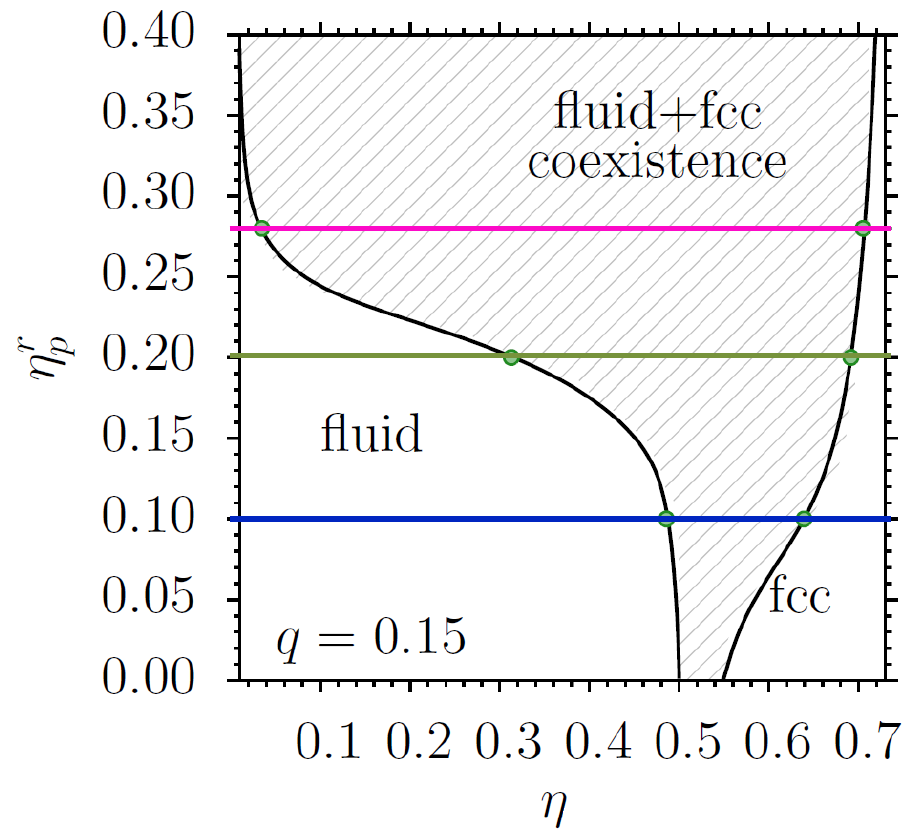}
\caption{\label{fig:pd}Schematic phase diagram of the softEffAO model in the plane of variables $\eta_p^r$ and $\eta$. The shaded region shows the fluid-crystal coexistence regions limited by the curve $\eta_f(\eta_p^r)$ on the left and by the curve $\eta_m(\eta_p^r)$ on the right. The three choices $\eta_p^r=0.1$, 0.2 and 0.28 that are extensively studied here are highlighted by horizontal straight lines.}
\end{figure}

\begin{figure*}
\centering
\subfigure[]{\includegraphics[width=0.45\textwidth]{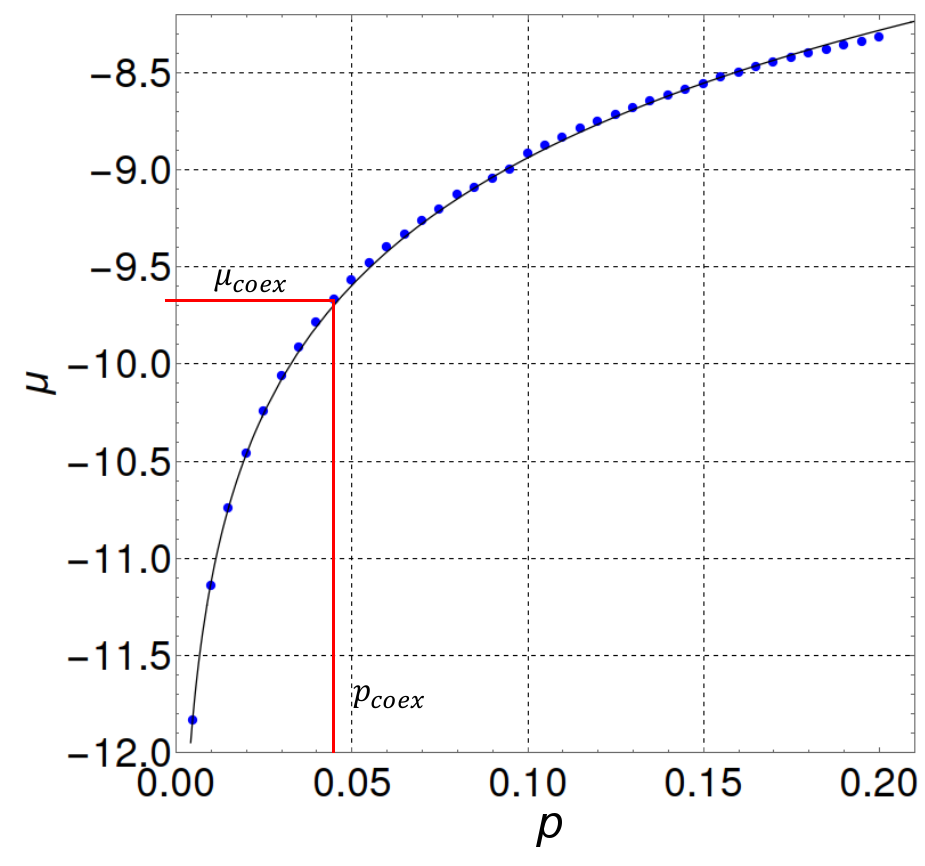}}
\subfigure[]{\includegraphics[width=0.45\textwidth]{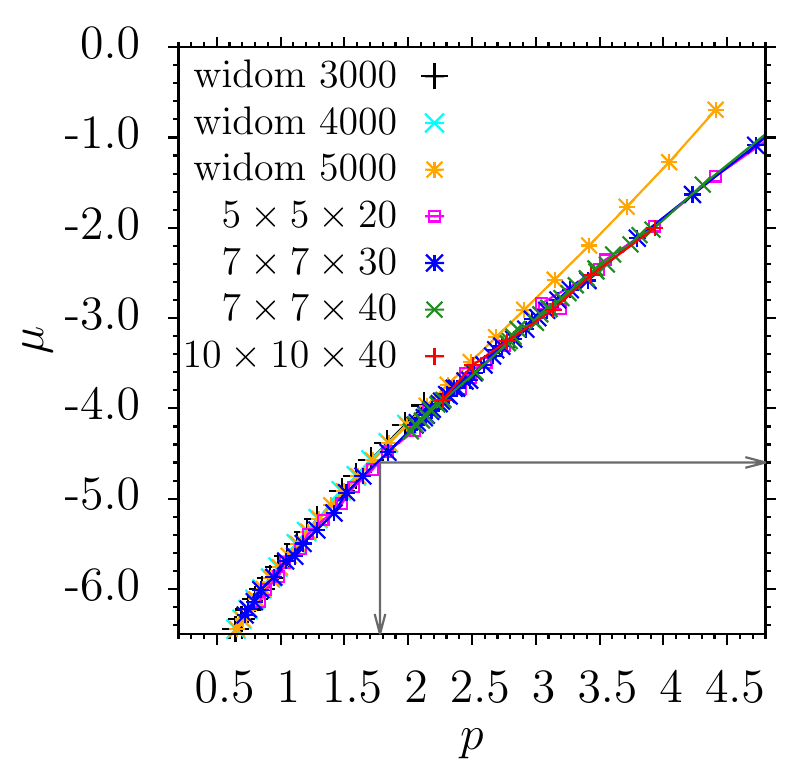}}
\caption{\label{fig:widom}a) Chemical potential of the colloids at $\eta_p^r=0.28$ plotted as a function of the pressure. Error bars are at most the size of the symbols. As expected for the regime where the colloidal suspension resembles an ideal gas, a logarithmic variation applies, $\mu=\mu_0+a\ln(p)$, with $\mu_0=-6.76\pm0.01$, $a=0.95\pm0.01$ (curve). The vertical straight line indicates the estimate for the coexistence pressure, the horizontal straight line indicates $\mu_{coex}$. The estimate for $p_{coex}$ shown here is taken from the interface velocity method, $p_{coex}\approx0.046\pm0.002$ and hence $\mu_{coex}\approx-9.7\pm0.1$. The chemical potential is measured in units of $k_BT$ with $k_BT=1$.\\
b) Chemical potential of the colloids at $\eta_p^r=0.20$ plotted as a function of the pressure. Various symbols show data from the Widom particle insertion method (using a $L\times L\times L$ box in the $NpT$ ensemble with $N=3000$, 4000, and 5000 colloid particles, as indicated) and from the repulsive wall potential mehod for 4 choices of linear dimensions $L$ and $L_z$, as indicated. The equation of state (Fig. \ref{fig:eos}) $p=p_l(\eta)$ was used to convert the data for $\mu$ as a function of $\eta=\eta_{bulk}$ (cf. text) into data for $\mu$ vs $p$. Arrows indicate the estimate for $p_{coex}$ (from the interface velocity method, see Sec. \ref{sec:bulkphase}) and the corresponding estimate for $\mu_{coex}$.}
\end{figure*}

The Asakura-Oosawa (AO) model \cite{57,58,59,60,61} describes colloids as hard spheres of diameter $\sigma_c$ and polymers as soft spheres (which may overlap each other with zero energy cost) of diameter $\sigma_p$, while overlap of colloids and polymers is strictly forbidden. Due to the ideal gas-like behavior of the polymers, one can integrate out their degrees of freedom exactly, thereby generating an effective colloid-colloid attraction (depletion potential). For a size ratio of $q=\sigma_p/\sigma_c<q^*=2/\sqrt{3}-1\cong 0.1547$ \cite{62} this effective potential is a simple pairwise interaction,
\begin{align}
U_{\text{eff}}(\sigma_c<r&<\sigma_c+\sigma_p)=\nonumber\\
\frac{\pi}{6}k_BT\sigma_p^3\frac{z_p}{(1+\frac{1}{q})^3}&\left[1-\frac{3\frac{r}{\sigma_c}}{2(1+q)}+\frac{\left(\frac{r}{\sigma_c}\right)^3}{2(1+q)^3}\right]~,
\end{align}
while $U_{\text{eff}}(r<\sigma_c)=\infty$ and $U_{\text{eff}}(r\geq\sigma_c+\sigma_p)=0$. Here $r$ is the distance between the centers of mass of the colloids, and $z_p=\exp(\mu_p/k_BT)$. To avoid the hard core singularity at $r=\sigma_c$, we rather add another repulsive potential to Eq. (14),
\begin{align}
U_{\text{rep}}(&r)=4[\left(\frac{b\sigma_c}{r-\epsilon\sigma_c}\right)^{12}+\left(\frac{b\sigma_c}{r-\epsilon\sigma_c}\right)^6\nonumber\\
- &\left( \frac{b \sigma_c}{\sigma_c+q-\epsilon\sigma_c} \right)^{12} - \left( \frac{b \sigma_c}{\sigma_c+q-\epsilon \sigma_c} \right)^{6}]~,
\end{align}
which is a smooth function, diverging at $r=\epsilon\sigma_c$ and vanishing at $r=\sigma_c+q$. Useful choices for the constants $b$ and $\epsilon$ are $b=0.01$, $\epsilon=0.98857$ (of course, other choices of these constants or other choices of a smooth function replacing the hard core potential would equally be possible). Fig. \ref{fig:softeffAO} shows the resulting potential $U(r)=U_{\text{eff}}(r)+U_{\text{rep}}(r)$ of this SoftEffAO model for three choices of the parameter $\eta_p^r$, commonly denoted as ``polymer reservoir packing fraction",
\begin{equation}
\eta_p^r=\frac{\pi\sigma_p^3}{6}e^{\frac{\mu_p}{k_BT}}~,
\end{equation}
for the size ratio $q=0.15$. 
Note that a solution of the polymers at concentration $\rho_p$ with no colloids present would have a polymer packing fraction of $\eta_p=(\pi\sigma_p^3/6)\rho_p=\eta_p^r$, of course; so variation of the polymer concentration in the colloidal suspension allows to control the strength of the colloid-colloid attraction. The fact that the range of this potenial, Eqs. (14), (15) is extremely short (and unlike the Lennard-Jones potential no artificial truncation is required) ensures a fast performance of the Monte Carlo simulation code.\\
To study the pure fluid and crystal phases of the system, it is most convenient to carry out standard $NpT$ Monte Carlo simulations, where apart from particle displacements in the cubic simulation box also volume changes are carried out \cite{69,70,71}. It was found that using $N=4000$ finite size effects are completely negligible, and hence the observed volume $\left<V\right>$ at a given pressure $p$ and inverse temperature-like variable yields the same equation of state as for a calculation where the pressure $\left<p\right>$ is sampled in the canonical $NVT$ ensemble. It is common to convert the density $\rho=N/V=1/v$, where $v$ is the volume per particle, to the colloid packing fraction $\eta$,
\begin{equation}
\eta=\frac{\pi}{6}\sigma_c^3\frac{N}{V}~.
\end{equation}
Fig. \ref{fig:eos} shows the equation of state found by these $NpT$ simulations for four choices of $\eta_p^r$, in the $\eta-p$ plane.
Ideally the fluid branch $p_l(\eta)$ should end at $p_{coex}=p_l(\eta_f)$ and exactly at the same pressure $p_{coex}=p_c(\eta_m)$ the crystal branch should start, so both phases together should exist only for the single pressure $p_{coex}$. However, the data reveal considerable hysteresis: there is a rather extended pressure interval $\Delta p$ aroud $p_{coex}$, such that for pressures from this interval both phases are found. For a pressure $p\neq p_{coex}$ from this interval only one phase is stable, the other being metastable. In fact, the estimation of the equilibrium coexistence conditions ($p_{coex}$, $\eta_f$, $\eta_m$) is a delicate and nontrivial task, and will be discussed in the next section. When this task is solved, the phase diagram can be drawn as a diagram of variables $\eta_p^r$ and $\eta$ that clearly reveals the fluid + fcc crystal two-phase coexistence region (Fig. \ref{fig:pd}).
From Fig. \ref{fig:pd} it is evident that the character of two-phase coexistence gradually changes from liquid/solid at small $\eta_p^r$, where $\eta_f$ is only a bit smaller than $\eta_m$, to gas/solid for large $\eta_p^r$, when $\eta_f\rightarrow 0$ while $\eta_m$ almost corresponds to close packing of hard spheres. The possibility that $\eta_f$ changes continuously from large values (typical for crystallization of melts) to very small values (typical for crystallization from the vapor phase) is a particular advantage of the present model which lacks a vapor to liquid transition and hence no triple point occurs; for the Lennard-Jones model, on the other hand, $\eta_f^{lc}$ for the liquid to crystal transition can decrease only a little with increasing temperature $T$, until the value $\eta_f^{lc}(T_t)$ at the temperature of the triple point is reached, and then vapor to crystal nucleation starts at $\eta_f^{vc}(T_t)\ll\eta_f^{lc}(T_t)$; thus there is a wide regime of values for $\eta_f$ for which no crystallization/melting transition can be studied.\\
In Sec.$\,$\ref{sec:intro} it was already pointed out that a particularly useful quantity to interpret simulations of phase coexistence is the chemical potential $\mu$. For the choice $\eta_p^r=0.28$ the packing fractions at the liquid branch are small and then the Widom particle insertion method \cite{72} can be straightforwardly applied and yields accurate results (Fig. \ref{fig:widom}).
However, for $\eta_p^r=0.1$ and $0.2$ the Widom particle insertion method cannot be applied straightforwardly. For $\eta_p^r=0.1$ the packing fractions of interest are near $\eta_f=0.494$ and there the success rate of particle insertions simply is too small. Therefore Statt \textit{et al.} \cite{63,64,73} proposed to use a system in a $L\times L\times L_z$ geometry with $L_z\gg L$, where near $z=0$ a soft repulsive wall potential $V_{\text{rep}}(z)$ is applied, which has a range $\sigma_w$ of about two colloid diameters. For the opposite wall a short-range strongly repulsive wall potential is chosen, while in $x$ and $y$ directions, PBC are still applied. Due to these walls at $z=0$ and $z=L_z$, a nonuniform packing fraction profile $\eta(z)$ develops in the system, but for large enough $L_z$ (e.g. $L_z=30$) this profile develops a strictly constant part $\eta(z)=\eta_{bulk}$ in the center of the system, e.g. for $z_{min}\leq z\leq z_{max}$ we can compute $\eta_{bulk}$ from an average over the profile accurately, $\eta_{bulk}=(z_{max}-z_{min})^{-1}\int_{z_{min}}^{z_{max}}\eta(z)dz$. Note that $\eta_{bulk}$ differs from the average packing fraction $\eta$ in the system due to wall excess contributions, of course \cite{63,73}. On the other hand, general principles of statistical mechanics \cite{39} imply that in equilibrium the chemical potential of such a system with inhomogeneous density is strictly constant throughout the system, $\mu(z)\equiv\mu$, even for $z$ close to the repulsive wall where $\eta(z)$ is much smaller than $\eta_{bulk}$. Measuring $\mu(z)$ by the particle insertion method for $z<\sigma_w$ this fact has been verified in the simulation, and although one uses only a small part $\Delta z\ll L_z$ of the total simulated system for the estimation of $\mu$, a very good accuracy can practically be established \cite{63,64,73}. We also have verified that for $\eta_p^r=0.2$ in the range $p\leq 2.5$, which corresponds to $\eta_c\leq 0.39$, the agreement between both methods to estimate $\mu(p)$ is very good. For larger pressures, however, the Widom particle insertion method leads to a systematic overestimation of $\mu$; we attribute this problem to the fact that for large $\eta$ (and hence large $p$) the distribution of the chemical potential that is sampled by the insertion method gets very asymmetric around its maximum. Analogous data for $\mu$ vs $p$ in the case $\eta_p^r=0.1$ can be found in \cite{63,64}. For the estimation of the nucleation barrier $\Delta F^*=\frac{1}{2}(p_c-p_l)V_n^*$ $\{$Eq. (6)$\}$ we shall need the pressure $p_l$ of the fluid surrounding the crystal nucleus (of volume $V_n^*$) as well as the associated crystal pressure $p_c$. However, this is not easily accessible directly, and hence a thermodynamic integration procedure has been applied. We use the relation for the chemical potential of the crystal
\begin{equation}
\mu_c(p_c)=\mu_{coex}+\int_{p_{coex}}^{p_c}dp'\frac{\pi}{6\eta_c(p')}~,
\end{equation} 
making use of the fact that $\mu_c(p_c)=\mu_l(p_l)$, which follows from Eqs. (11), (12). For large enough nuclei it turns out that $\eta_c(p_c)$ differs only very little from $\eta_c(p_{coex})=\eta_m$, and then Eq. (18) gets simplified as 
\begin{equation}
\mu_c(p_c)=\mu_{coex}+\frac{\pi}{6\eta_m}(p_c-p_{coex})~,
\end{equation}
and hence the knowledge of $p_{coex}$, $\mu_{coex}$ together with $V_n^*$, $p_l$, $\mu_l(p_l)$ will suffice to allow the computation of the desired free energy barrier $\Delta F^*$. Hence we proceed to discuss the estimation of phase coexistence conditions in the bulk.
\section{Characterizing bulk two-phase coexistence\label{sec:bulkphase}}
In the discussion of Fig. \ref{fig:eos} we have already pointed out that in the $NpT$ simulations of bulk systems we find an extended range of pressures near $p_{coex}$ where either the fluid phase or the crystal phase is metastable, and during affordable lengths of the simulation run no transition to the stable phase occurs. Thus these data cannot be used to estimate $p_{coex}$ and hence $\eta_f$ and $\eta_m$. 
\begin{figure}
\centering
\includegraphics[width=0.45\textwidth]{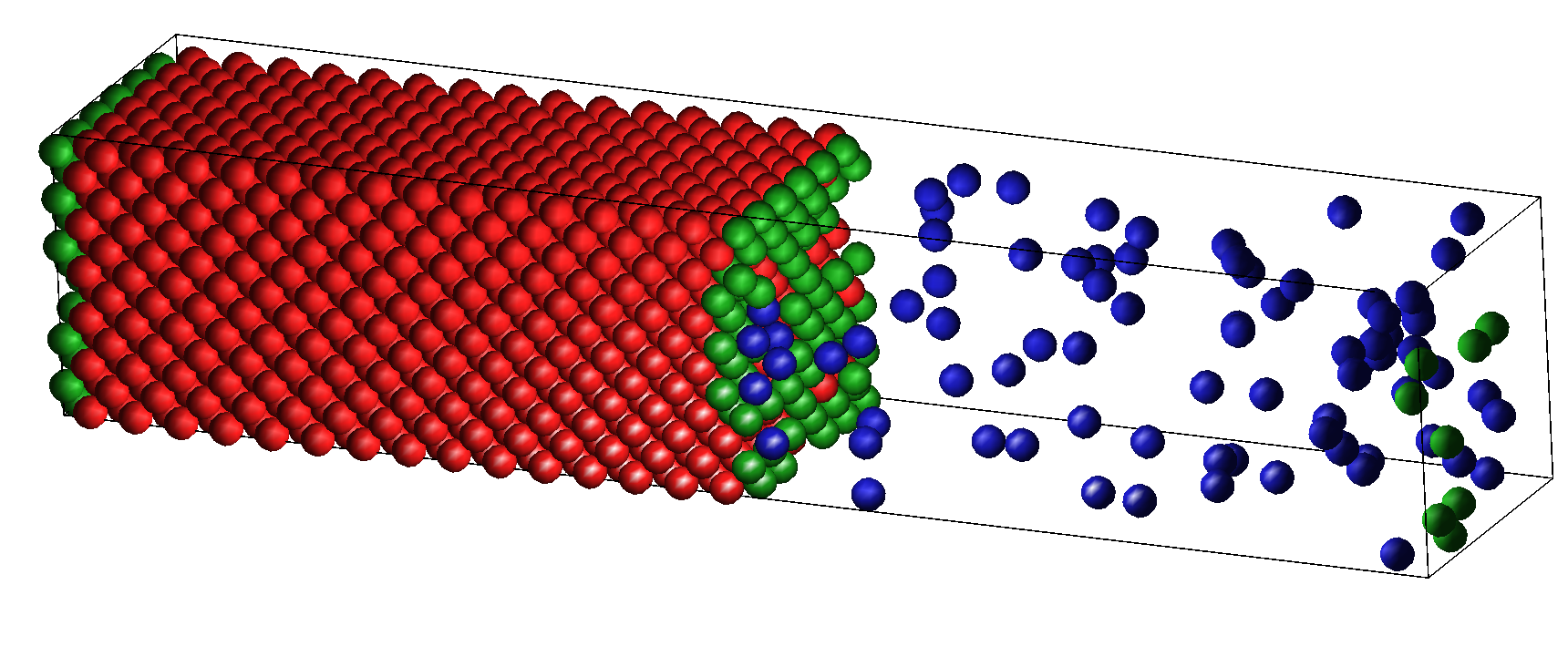}
\caption{\label{fig:interfacevel}Equilibrated configuration of a fcc crystal (taking a volume $L\times L\times L_z/2$) coexisting with fluid (also taking a volume $L\times L\times L_z/2$) for $\eta_p^r=0.28$ and choosing $n=6$ at a pressure $p_z=p=p_{coex}$. Particles in the crystal that have 12 nearest neighbors in the bulk and at least 9 neighbors in the last layer before the interface plane are colored red, particles in the fluid phase (which has less than 6 neighbors here) are shown in blue, while particles with 6-8 neighbors in the interfacial region are displayed in green.}
\end{figure}
\begin{figure*}
\centering
\subfigure[]{\includegraphics[width=0.45\textwidth]{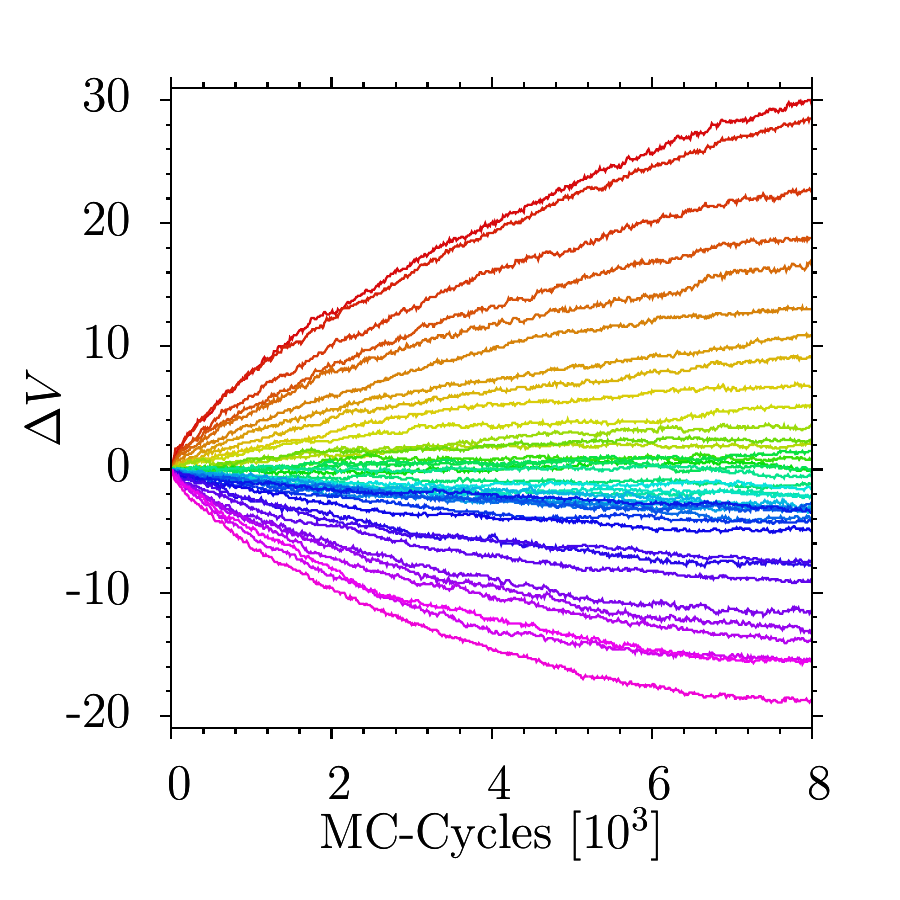}}
\subfigure[]{\includegraphics[width=0.41\textwidth]{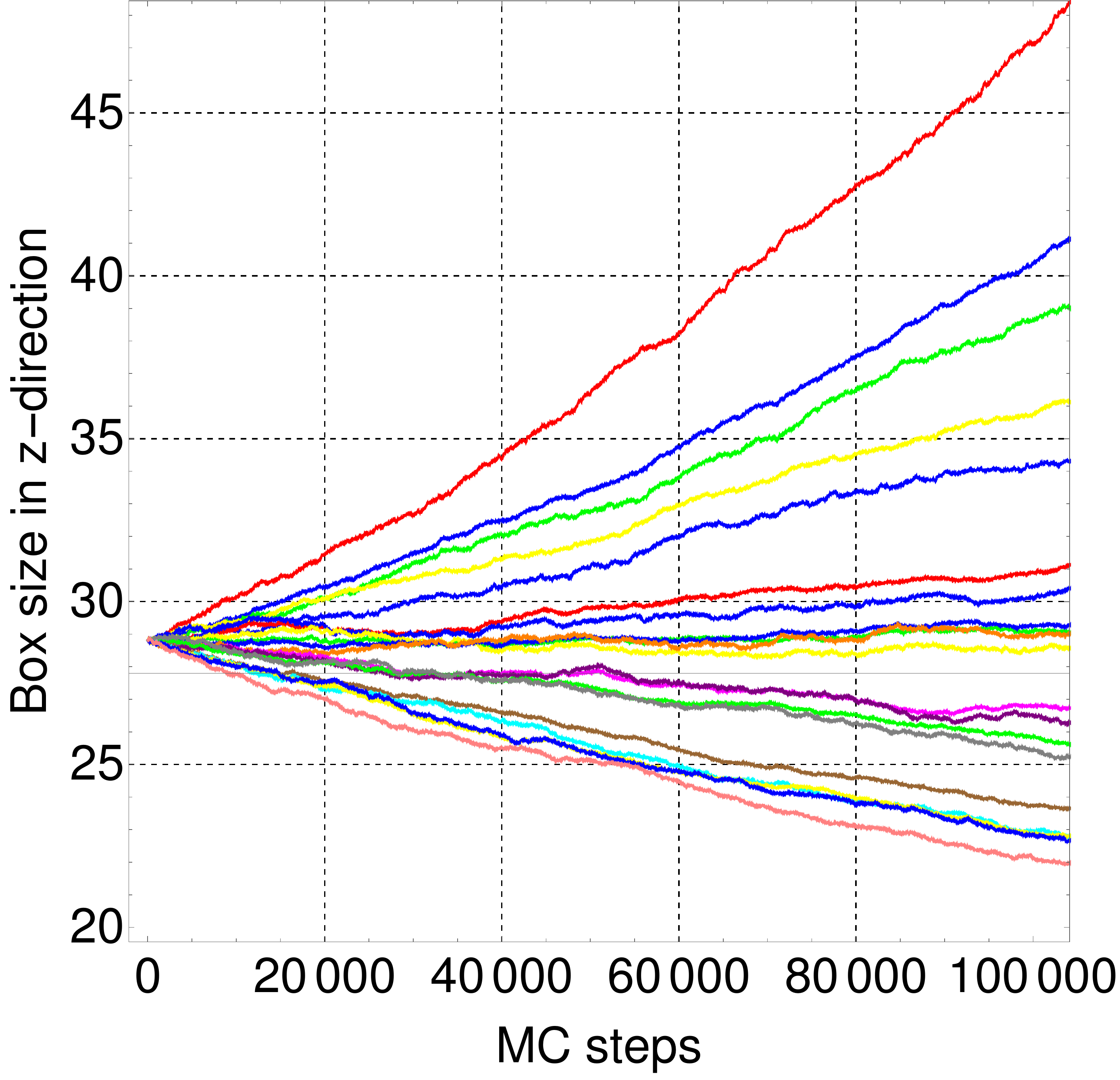}}
\caption{\label{fig:dVdT}a) Volume changes $\Delta V(t)$ vs. time $t$ (in units of $10^3$ MC cycles) for the case $\eta_p^r=0.2$ and $n=10$, varying the pressure $p$ from $p=0.6$ (red, top) to $p=3.0$ (magenta, bottom). For $p$ near $p_{coex}$ the slope d$\Delta V(t)/$d$t\approx 0$.\\b) Box linear dimension $L_z$ vs. time $t$ for the case $\eta_p^r=0.28$ and $n=4$, from $p=0.005$ (red, top) to $p=0.1$ (pink, bottom curve). For $p$ near $p_{coex}$ the slope d$L_z(t)/$d$t\approx 0$.}
\end{figure*}
To overcome this problem, we apply the approach of Zykova-Timan \textit{et al.} \cite{24,25} setting up a two-phase configuration of coexisting fcc crystal and fluid separated by two interfaces in an elongated box of linear dimensions $L\times L\times L_z$ with $L_z=5L$ with PBC throughout, simulated in the $Np_zT$ ensemble. 
The orientation of the crystal axes is chosen such that 2 interfaces of (100)-type occur, consistent with the geometry, and $L$ is chosen such that an integer number of lattice planes fits into the simulation box in $x$- and $y$- directions without misfit at the chosen pressure; the data for $\eta_c(p)$ at the crystal branches in Fig. \ref{fig:eos} are used to choose the lattice parameter and hence $L$ (ranging from $n=4$ to $n=12$ lattice cells in $x$- and $y$- directions) appropriately. Fig. \ref{fig:interfacevel} shows an example for $n=6$ and $\eta_p^r=0.28$, choosing both phases in one half of the box volume.
Note that in this case the fluid rather is a vapor-like phase, while for $\eta_p^r=0.1$ it rather is like a dense melt, and the quantity that is actually measured is the total box volume $V(t)$, or equivalently the linear dimension $L_z(t)$, as a function of simulation time t; note that after initialization the system is first equilibrated in the $NVT$ ensemble, choosing particle displacement Monte Carlo moves only, to allow that local equilibrium is established in the interfacial regions. So the origin $t=0$ of simulation time $t$ means the end of this $NVT$ equilibration period, when one switches to the $Np_zT$ simulation, now including volume changes by moves $L_z\rightarrow L_z\pm\Delta L_z$, as is standard \cite{69,70,71}. If $p=p_z<p_{coex}$, the crystal should be unstable and hence particles ``evaporate'' from the interfacial region into the fluid, and thus the volume of the system must increase, to maintain the proper (low) density of the fluid at the chosen pressure. On the other hand, if $p=p_z>p_{coex}$, the fluid is unstable, particles ``condense'' in the interfacial region, and the total volume shrinks. The actual observations of $V(t)$ or $L_z(t)$ are consistent with these expectations; of course, there are huge random fluctuations from run to run, and thus 50 to 120 runs were averaged in order to obtain meaningful accuracy (Fig. \ref{fig:dVdT}).
Of course, the inspection of the ``raw data'' for these time evolutions of crystal growth or shrinking only allows for a rather rough estimation of $p_{coex}$. Thus we tried to fit an interface velocity by fitting $V(t)$ or $L_z$ as a linear function over some time interval, but typically there is a systematic dependence of the resulting interface velocity on the size $\Delta t$ of this interval. 
\begin{figure}
\centering
\subfigure[]{\includegraphics[width=0.22\textwidth]{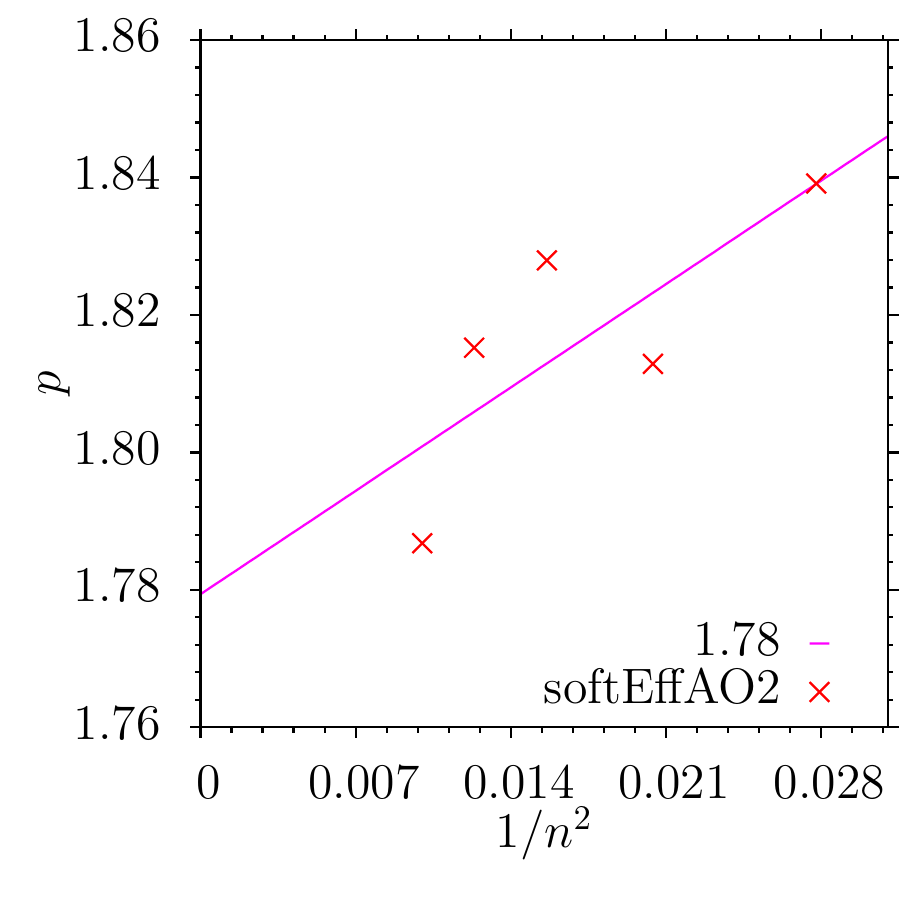}}
\subfigure[]{\includegraphics[width=0.23\textwidth]{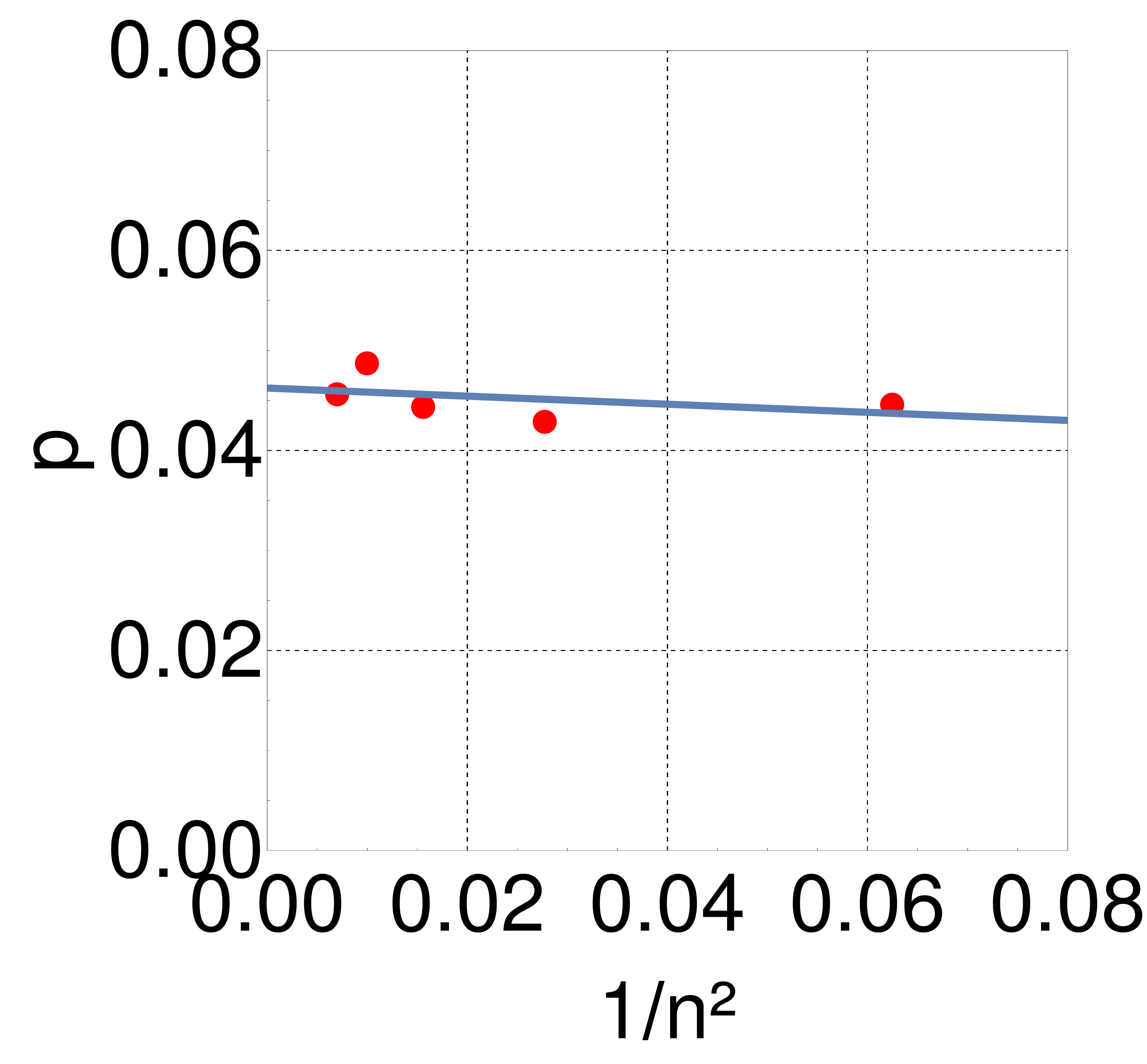}}
\caption{\label{fig:fsize}Estimates for $p_{coex}$ plotted vs $n^{-2}$ for $\eta_p^r=0.2$ (a) and $\eta_p^r=0.28$ (b). Straight line fits result in $p_{coex}=1.78\pm0.02$ (a) and $0.046\pm0.002$ (b).}
\end{figure}
\begin{figure*}
\centering
\subfigure[]{\includegraphics[width=0.55\textwidth]{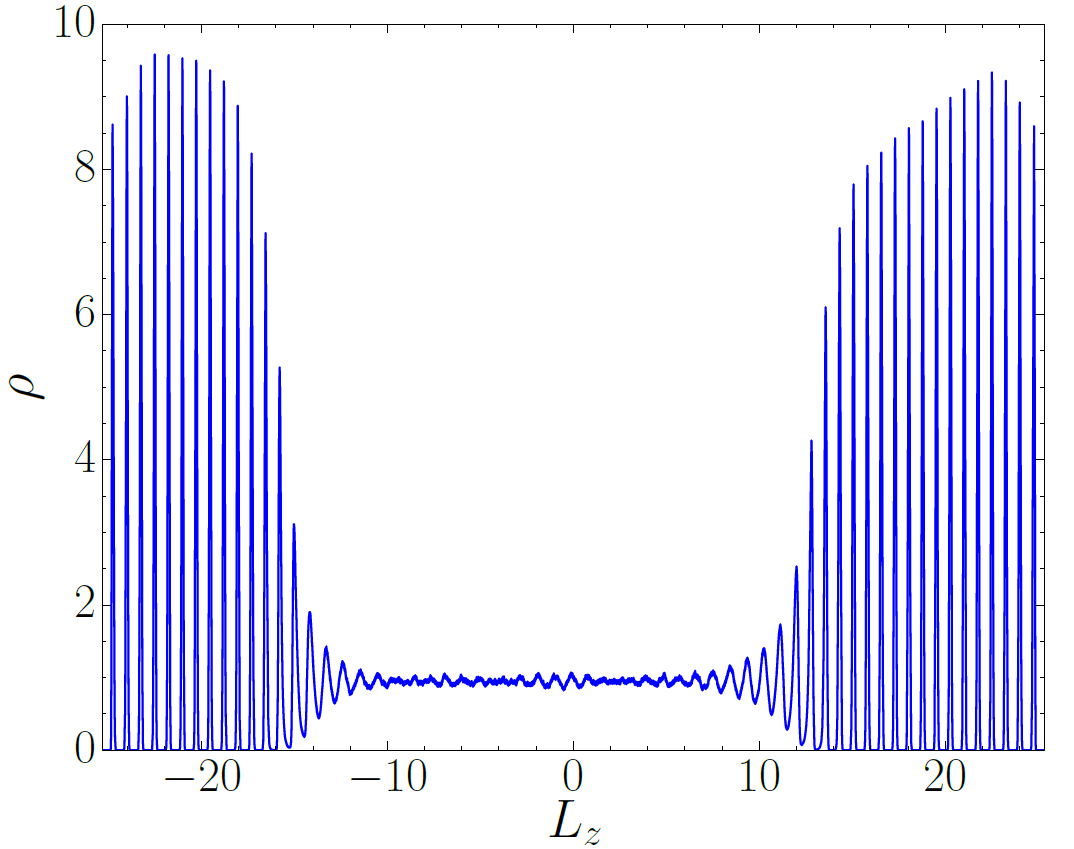}}
\subfigure[]{\includegraphics[width=0.38\textwidth]{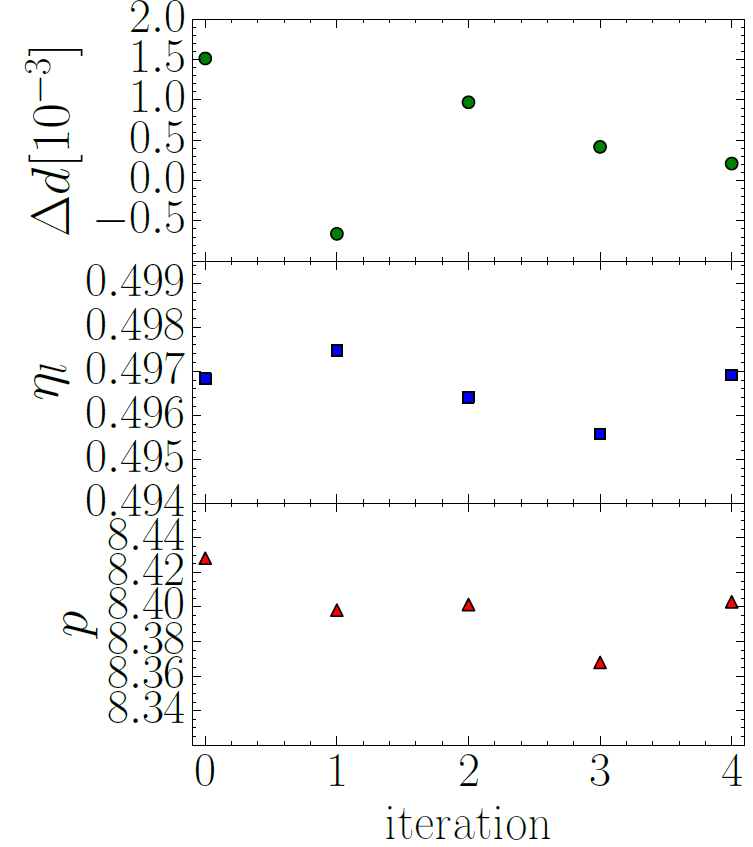}}
\caption{\label{fig:densmethod}(a) Density histogram of the softEffAO model with $\eta_p^r=0.1$ between two walls with distance $D=50.6$ in z direction. (b) Difference in lattice distances perpendicular and parallel to the wall $\Delta d=d_{x,y}-d_{z}$, packing fraction in the middle of the box, and
pressure in the fluid phase between the two crystal slabs as function
of number of iteration.}
\end{figure*}
\begin{figure*}
\centering
\subfigure[]{\includegraphics[width=0.31\textwidth]{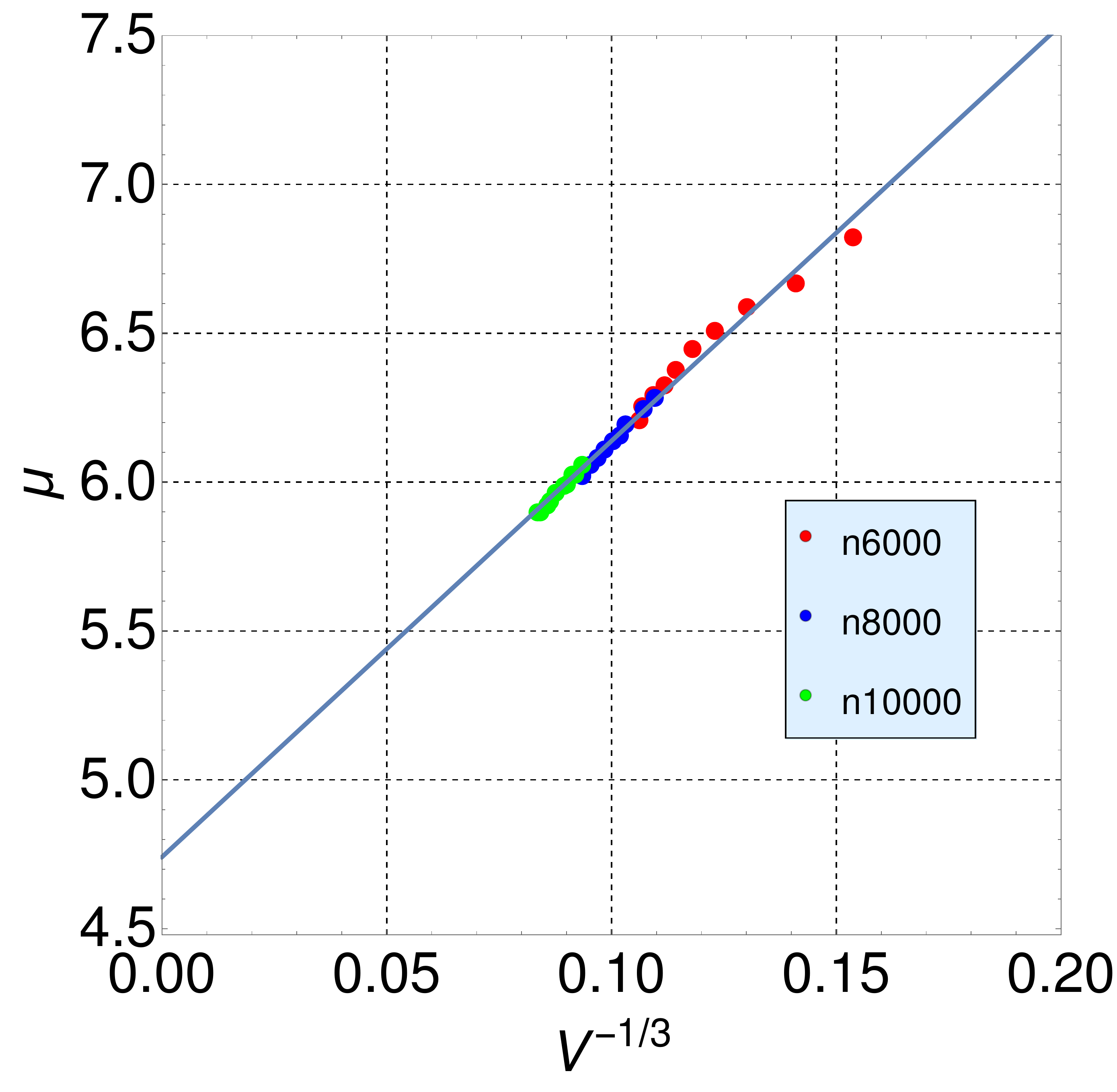}}
\subfigure[]{\includegraphics[width=0.31\textwidth]{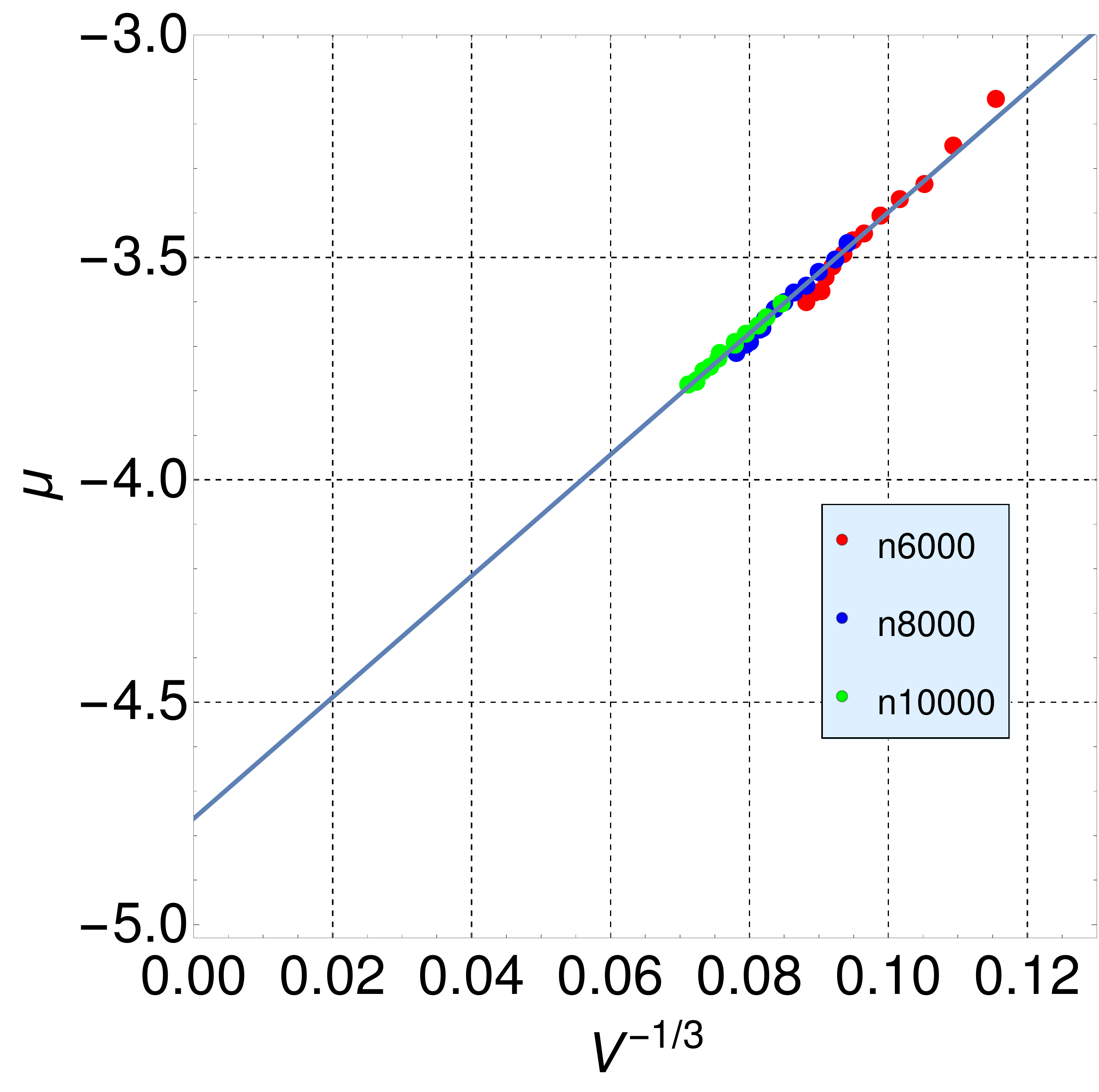}}
\subfigure[]{\includegraphics[width=0.31\textwidth]{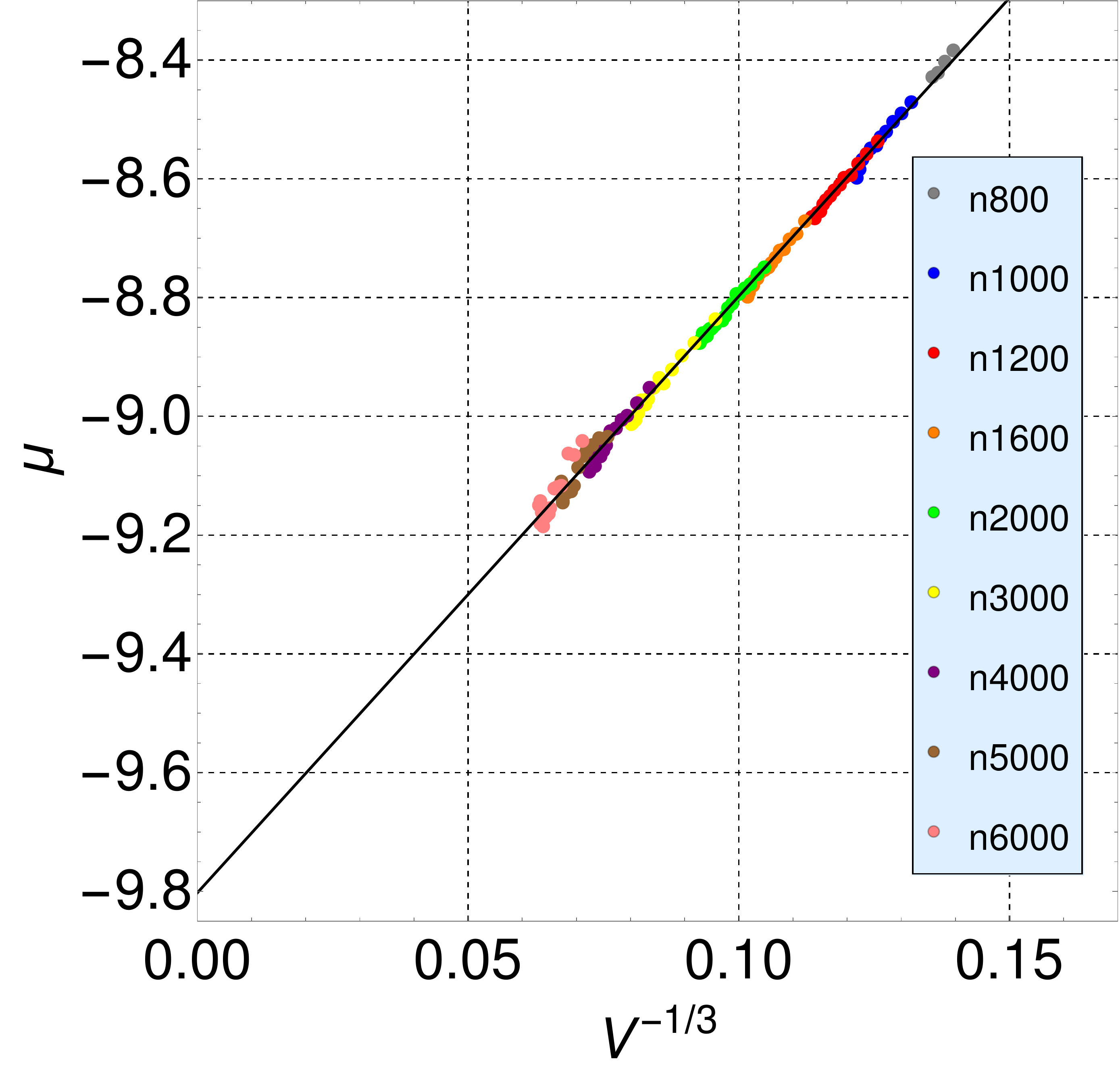}}
\caption{\label{fig:chvsvol}Chemical potential $\mu_l$ of the fluid surrounding a crystalline nucleus of volume $V_n^*$ plotted vs $(V_n^*)^{-1/3}$ for $\eta_p^r=0.1$ (a), 0.2 (b) and 0.28 (c). The straight line intercepts are $\mu_{coex}=4.74\pm0.03$ (a), $-4.77\pm0.02$ (b) and $-9.80\pm0.01$ (c). The corresponding results extracted via $p_{coex}$ (as obtained from the interface velocity method) from the equation of state were $4.74\pm0.04$ (a), $-4.60\pm 0.04$ (b) and $-9.67\pm 0.04$ (c). The data were based on simulations using $N=6000$, 8000 or 10000 particles in cases a) and b), and $N=800$ up to 6000 in case c) as indicated.}
\end{figure*}
Fortunately, the resulting estimate of $p_{coex}$ as the pressure where this interface velocity changes sign shows almost no dependance on this interval. However, there seems to be a significant finite size effect: thus our final estimates were obtained by carrying out an extrapolation versus $1/n^2$ towards the thermodynamic limit (Fig. \ref{fig:fsize}).
For $\eta_p^r=0.1$ (data shown in \cite{63,64} and not repeated here) the result was $p_{coex}=8.45\pm 0.04$, excluding the smallest value of n from the fit. Clearly the fluctuations in the data of Fig. \ref{fig:fsize} (and of \cite{63,64}) are large, and do not prove that the asymptotic regime of the extrapolation versus the inverse interfacial area really has been reached; thus a possible systematic error in these estimates for $p_{coex}$ cannot be excluded.\\
Clearly, it would be nice to have alternative methods to obtain $p_{coex}$. Deb \textit{et al.} \cite{55} suggested to use a $L\times L\times L_z$ geometry in the $NVT$ ensemble, with two planar repulsive walls at $z=0$ and $z=L_z$, and choose a packing fraction such that $\eta_f<\eta<\eta_m$. The system is initialized using an initial guess for $\eta_m$ (note that only when $p_{coex}$ has been determined are $\eta_f$ and $\eta_m$ accurately known!) to choose $L$ compatible with an integer number of lattice planes parallel to the $xy$-plane, and one chooses a few crystal layers parallel to the walls at $z=0$ and $z=L_z$, such that the remaining volume is compatible with the fluid phase at packing fraction $\eta_f$. The idea is that the system is self-regulating: if the packing fraction in the volume region available for the fluid is too high, particles will condense in the interfacial region, to make the crystalline layers a bit thicker, until $\eta_f$ in the fluid region would be established. However, if the initial guess for $\eta_m$ is not accurate enough, the crystal at $p_{coex}$ would be under strain, and hence this method would lead to a systematic error in the estimation of $p_{coex}$ and $\eta_f$ as well. Thus one needs to test a range of estimates for $\eta_m$ to find the choice where a $p_{coex}$ results for which the crystalline layers are undistorted. Of course, since not too few lattice planes parallel to the planar walls must be used at both walls, $L_z$ has to be large, and hence equilibration is rather slow. As a consequence, the accuracy reached by this method is not better than the method described in Figs. \ref{fig:interfacevel}-\ref{fig:fsize}, and yields significantly lower values for $p_{coex}$ (see Fig. \ref{fig:densmethod}).
Still another variant of phase coexistence simulation method due to Pedersen \cite{74} studies the interface between crystal and fluid under the action of a force that prevents its motion if $p\neq p_{coex}$, and the magnitude of the force is varied until one finds the pressure for which the interface does not move for zero force. However, we have not yet tried out this approach for the present model, but rather we have used an approach based on the phase coexistence of compact droplets with surrounding fluid, as will be described in the next section. Combining Eqs. (5), (19) with an analogous expansion for $p_l$
\begin{equation}
p_l\cong p_{coex}+\frac{6}{\pi}\eta_f(\mu_l-\mu_{coex})~,
\end{equation}
we find
\begin{equation}
\mu_l=\mu_{coex}+\frac{\pi}{9}\frac{A_W\bar{\gamma}}{\eta_m-\eta_f}(V_n^*)^{-1/3}~~,~~V_n^*~\rightarrow~\infty~.
\end{equation}
This relation suggests to plot the chemical potential $\mu_l$ of the fluid phase surrounding the crystal nucleus linearly vs. $(V_n^*)^{-1/3}$: if the data fall on a straight line that is independent of the total volume $V$ in Eq. (13), we can estimate $\mu_{coex}$ as an intercept of the straight line for $(V_n^*)^{-1/3}\rightarrow 0$. Fig. \ref{fig:chvsvol} shows that this recipe indeed works, and the accuracy seems to be competitive with the accuracy of the interface velocity method \cite{25,26}.
While the two methods do not agree within the error bars, resulting from the respective fits, we stress that these errors do not include systematic errors, e.g. due to higher order terms in Eq. (21) or the problem that in Fig. \ref{fig:fsize} not all data points fall in the regime of $n^{-2}$ variation.\\
In principle, one can try a linear expansion similar to Eq. (21) also for the pressure $p_l$ to find $p_{coex}$, and the packing fraction $\eta_l(p_l)$ to find $\eta_f$. When one does this, one finds for $\eta_p^r=0.1$ that $p_{coex}=8.47\pm0.03$ and $\eta_f=0.498\pm0.001$, close to the estimates extracted from the interface velocity method. For $\eta_p^r=0.2$, the corresponding results are $p_{coex}=1.59\pm0.02$ and $\eta_f=0.376\pm0.002$, while for $\eta_p^r=0.28$ the functions $p_l$ vs. $(V_n^*)^{-1/3}$ and $\eta_l$ vs. $(V_n^*)^{-1/3}$ exhibit pronounced curvature, and hence a linear extrapolation then does not work, while a nonlinear extrapolation certainly would be compatible with the results found from the equation of state when $\mu_{coex}$ is used.\\ As a result, we conclude that inaccurate knowledge of coexistence conditions is an important factor limiting the possibility to perform a high precision test of the predictions of homogeneous nucleation theory.
\section{Analysis of the equilibrium between the crystalline nucleus and surrounding fluid\label{sec:equilibrium}}
\begin{figure*}
\renewcommand{\arraystretch}{0.0}
		\setlength{\tabcolsep}{-1pt}
		\centering
		\begin{tabular}{ c c c c }
		\raisebox{0.1\height}{\includegraphics[width=0.23\textwidth]{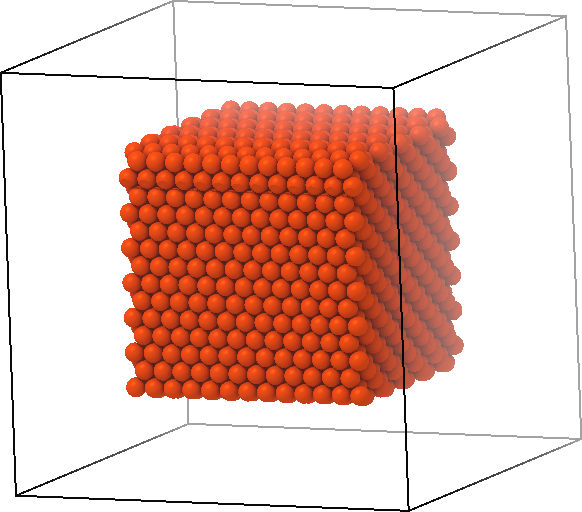}} & 
		\raisebox{0.1\height}{\includegraphics[width=0.23\textwidth]{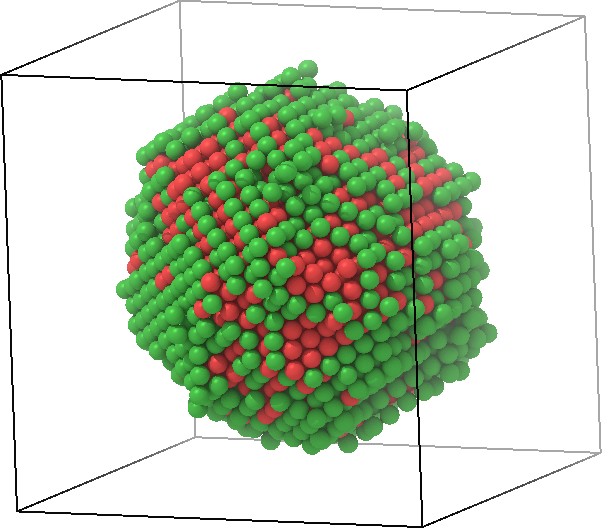}} & 
		\includegraphics[width=0.23\textwidth]{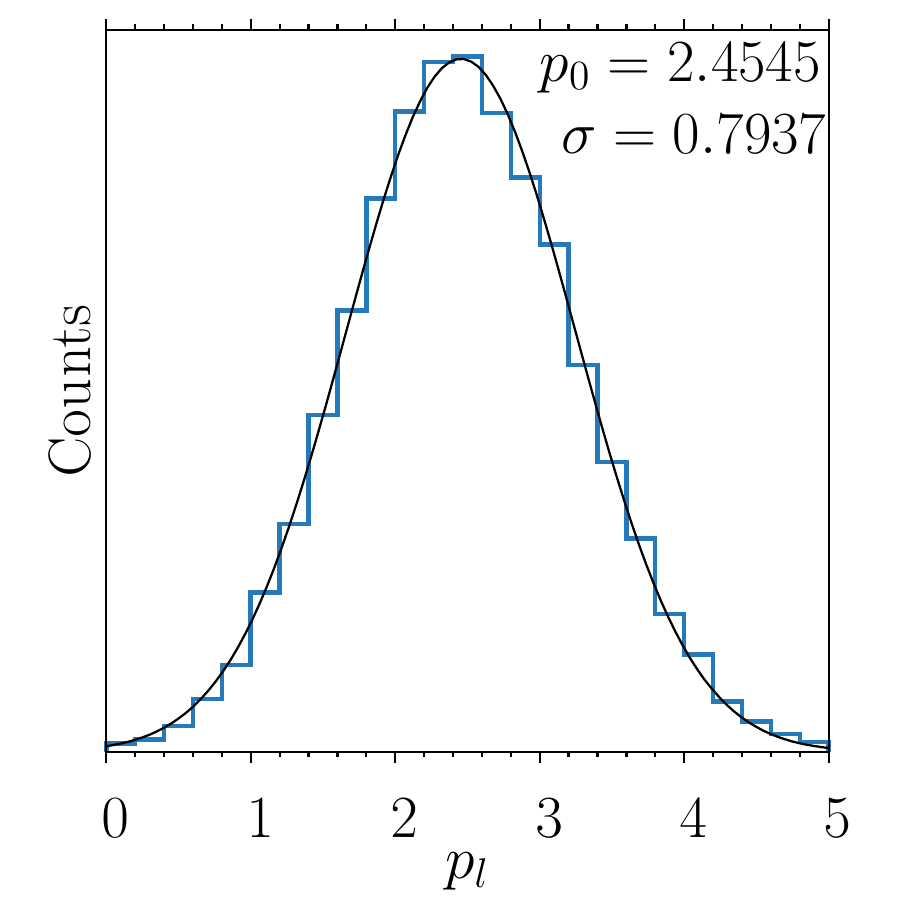} & 
		\includegraphics[width=0.23\textwidth]{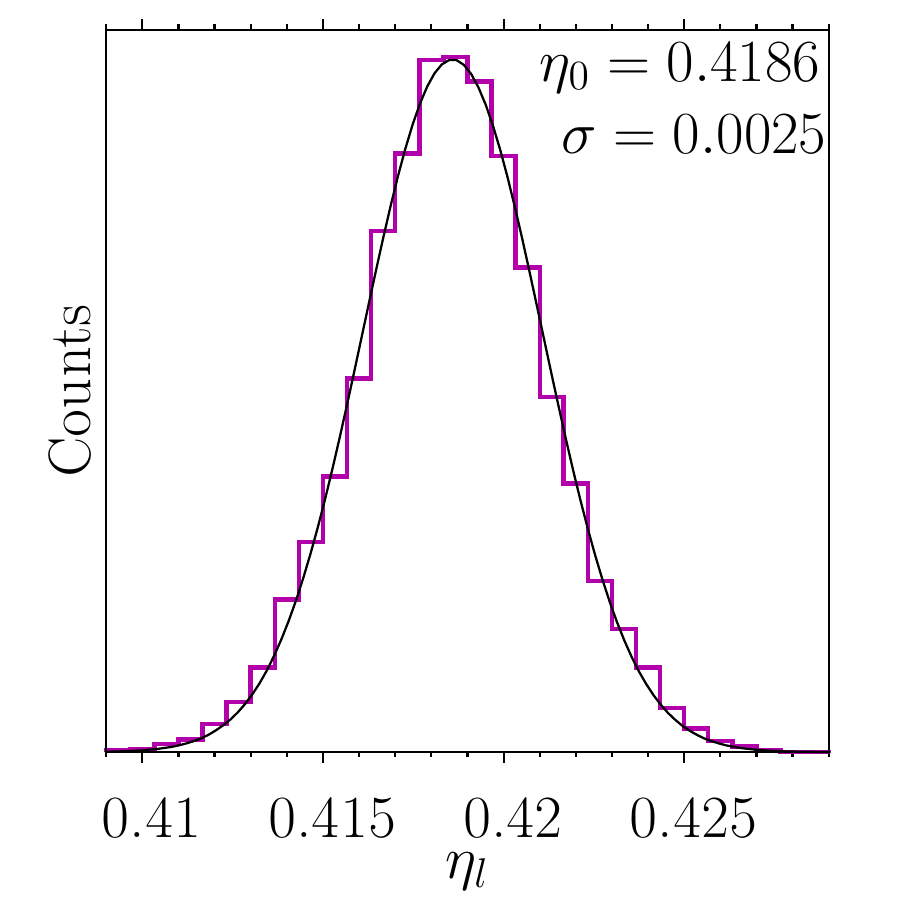} \\[0.1cm]
			(a) & (e) & (i) & (m) \\[0.45cm]
		\raisebox{0.1\height}{\includegraphics[width=0.23\textwidth]{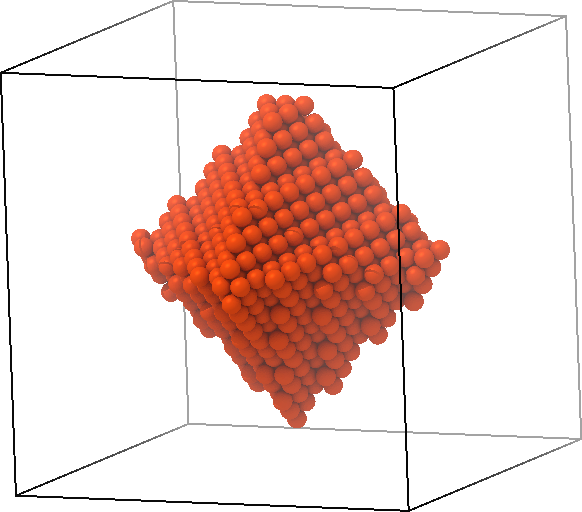}} & 
		\raisebox{0.1\height}{\includegraphics[width=0.23\textwidth]{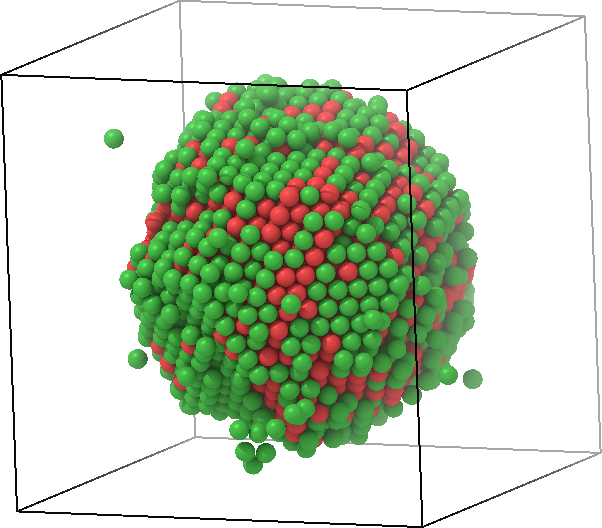}} & 
		\includegraphics[width=0.23\textwidth]{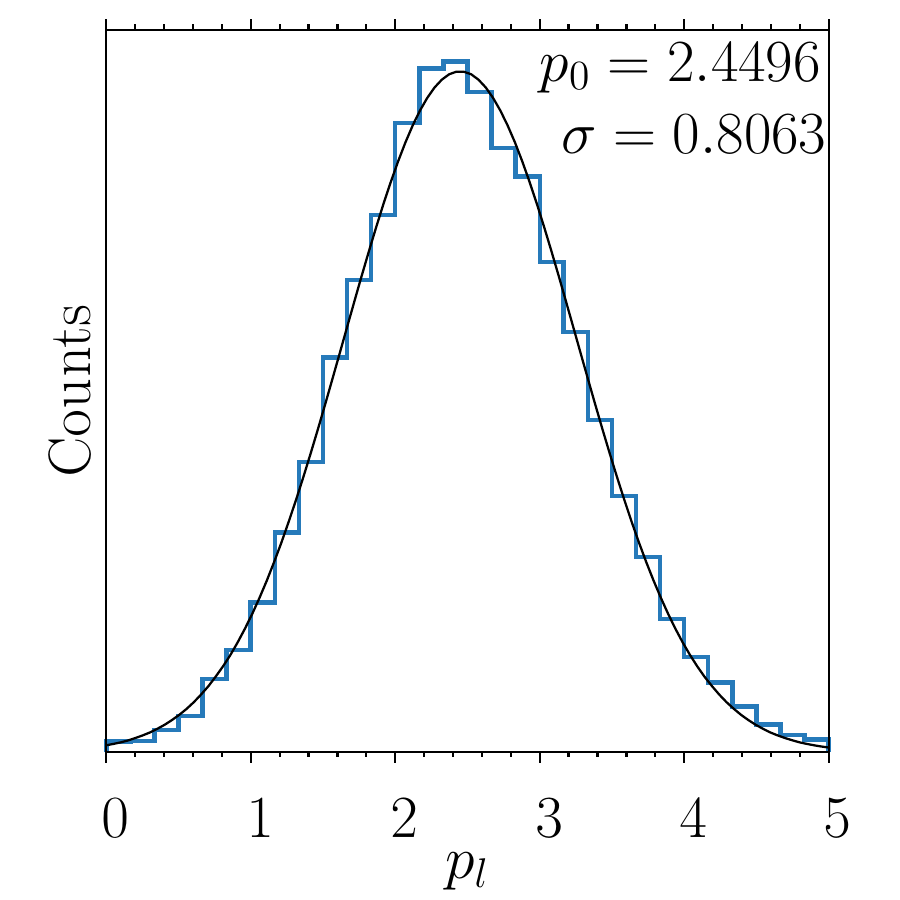} & 
		\includegraphics[width=0.23\textwidth]{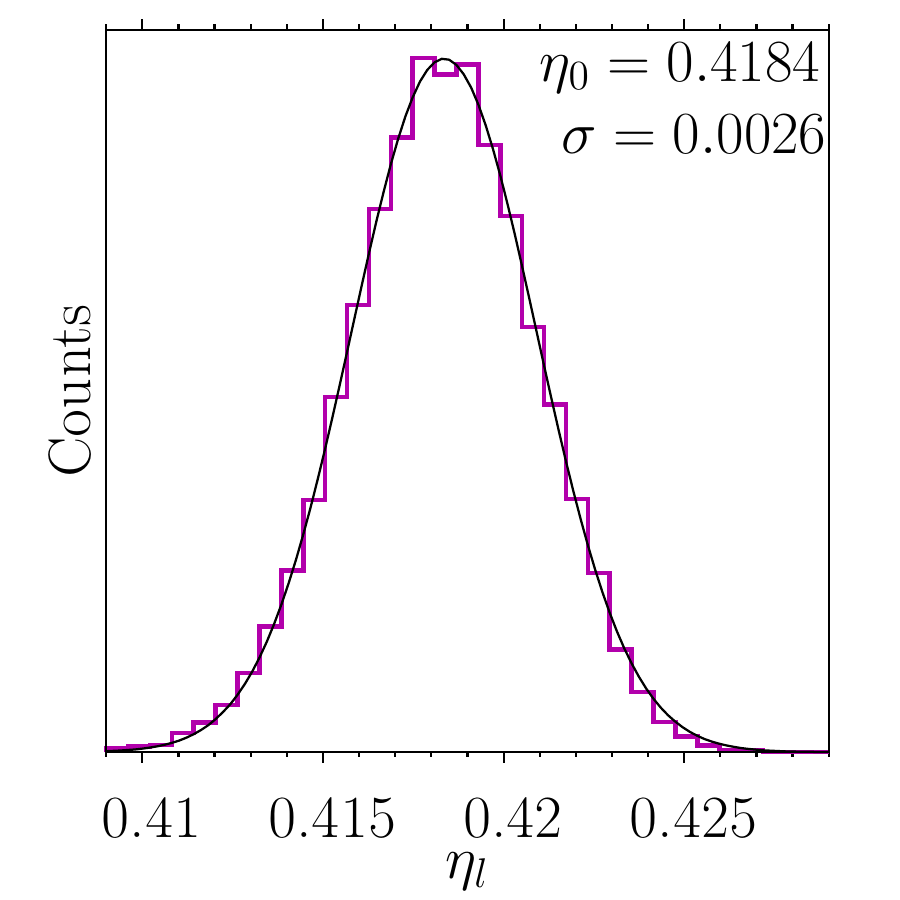}\\[0.1cm]
			(b) & (f) & (j) & (n) \\[0.45cm]
		\raisebox{0.1\height}{\includegraphics[width=0.23\textwidth]{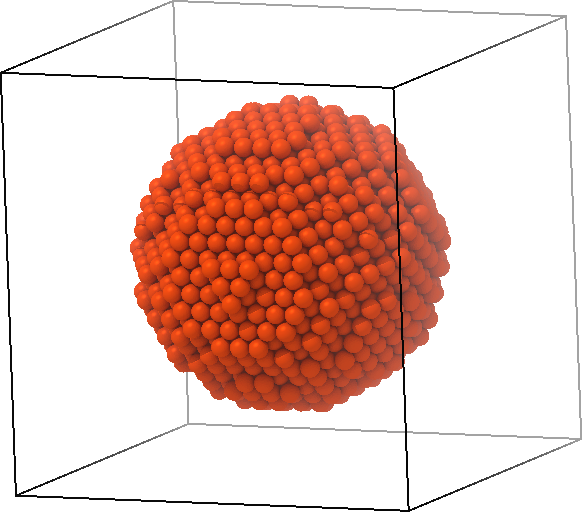}} & 
		\raisebox{0.1\height}{\includegraphics[width=0.23\textwidth]{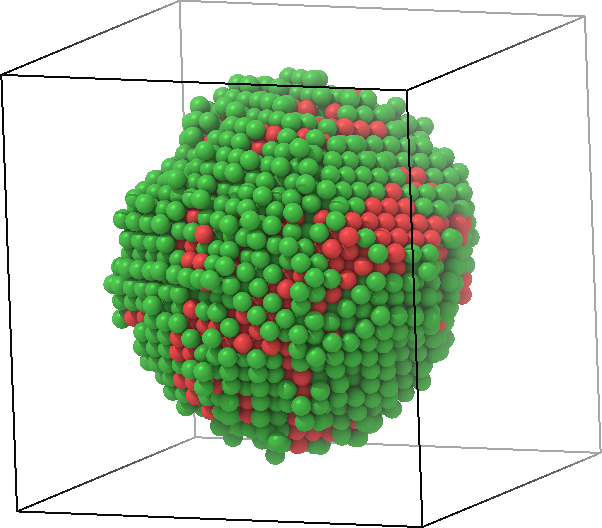}} & 
		\includegraphics[width=0.23\textwidth]{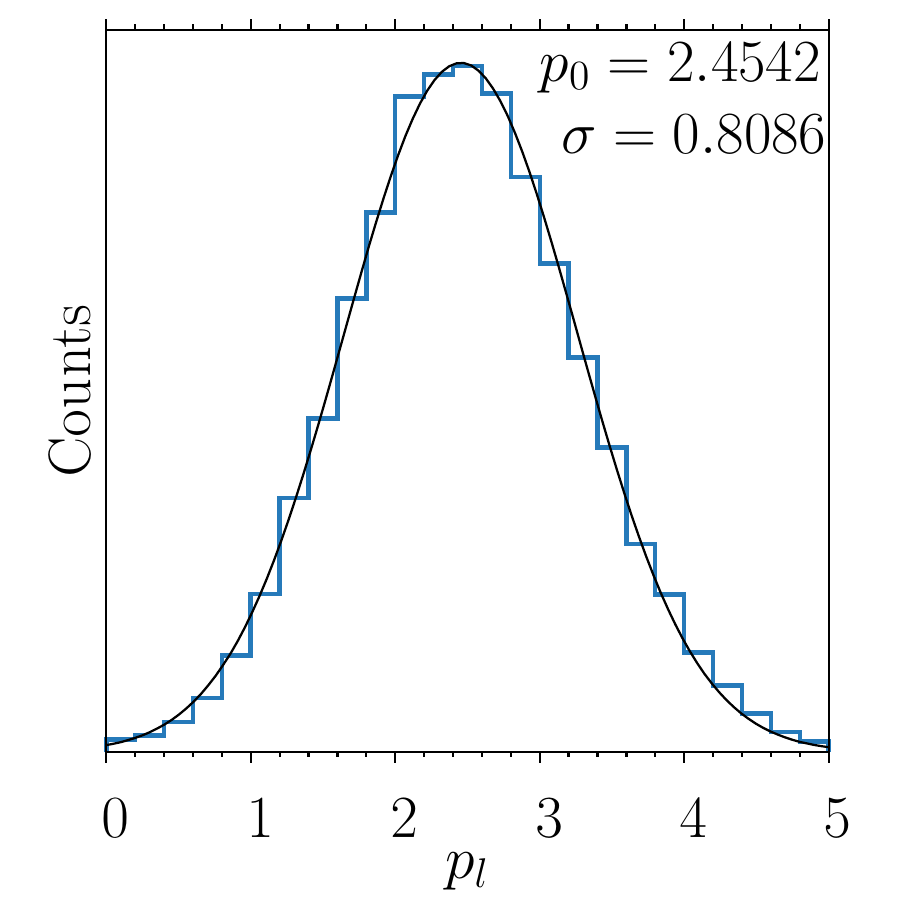} & 
		\includegraphics[width=0.23\textwidth]{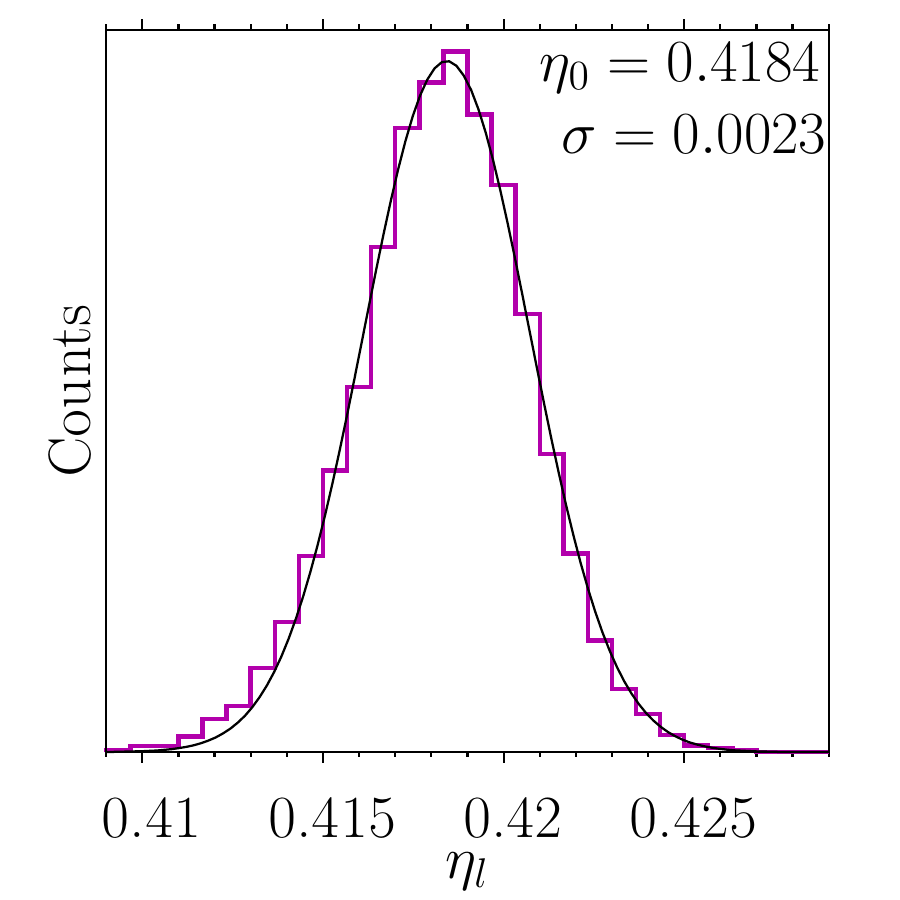} \\[0.1cm]
			(c) & (g) & (k) & (o) \\[0.45cm]
		\raisebox{0.1\height}{\includegraphics[width=0.23\textwidth]{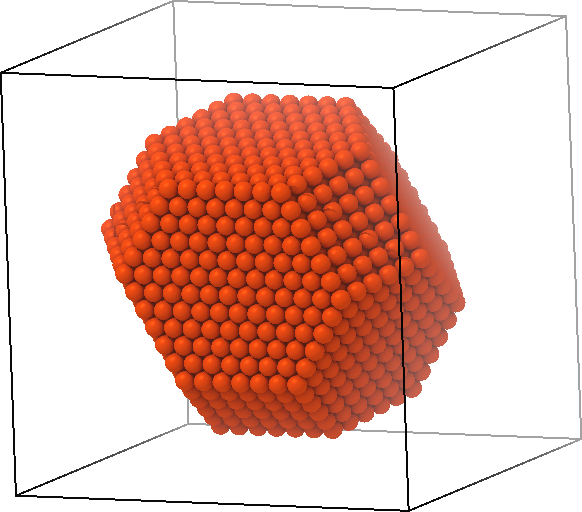}} & 
		\raisebox{0.1\height}{\includegraphics[width=0.23\textwidth]{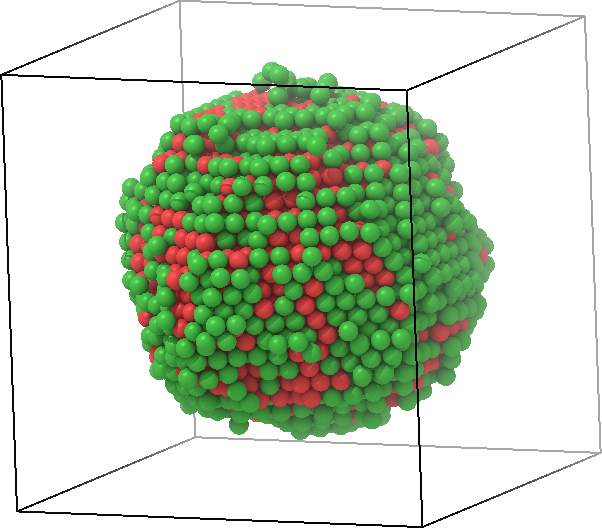}} & 
		\includegraphics[width=0.23\textwidth]{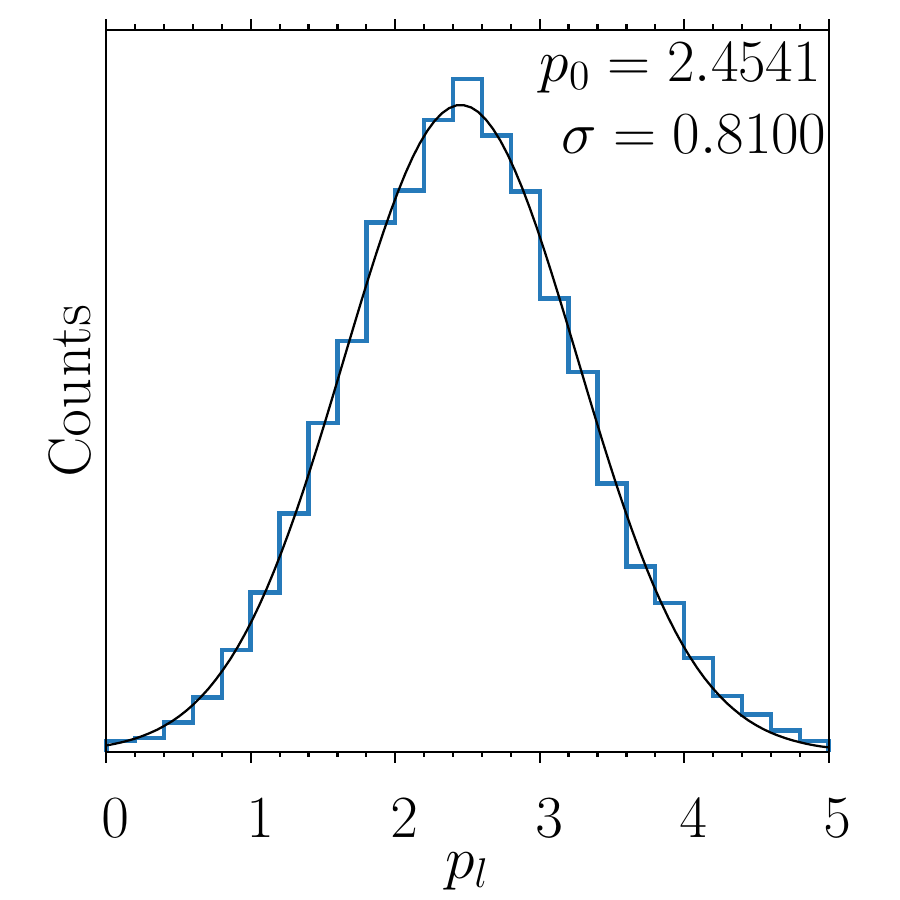} & 			\includegraphics[width=0.23\textwidth]{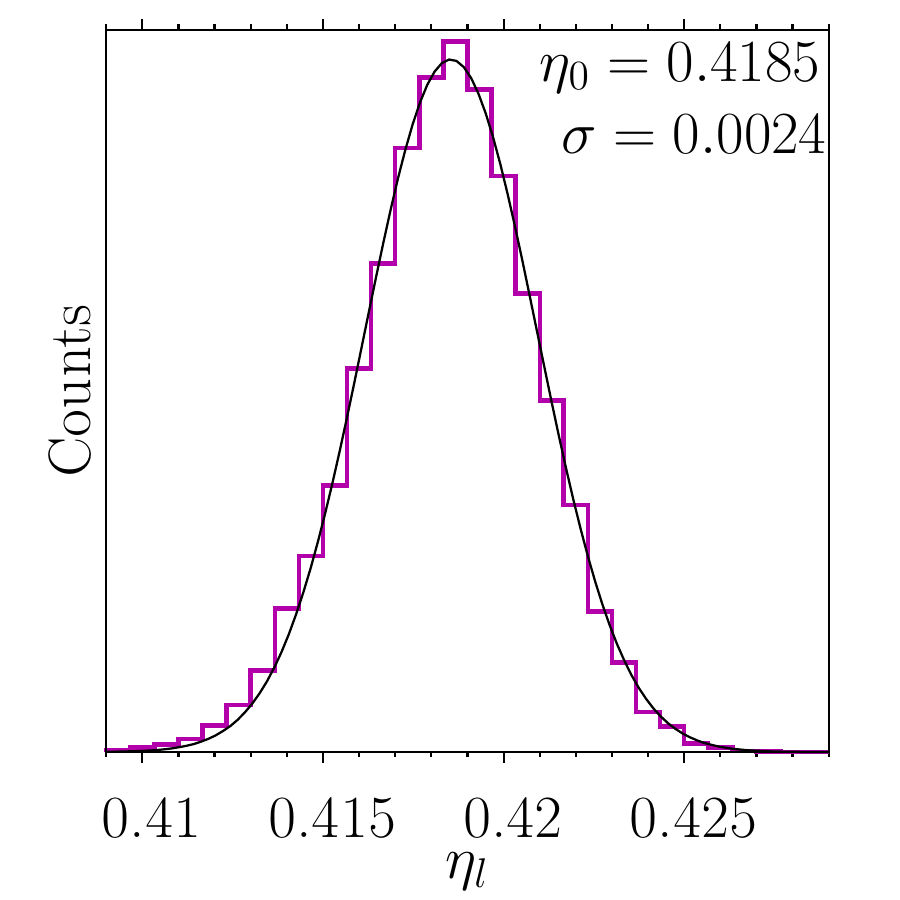} \\[0.1cm]
			(d) & (h) & (l)& (p)
		\end{tabular}
\caption{\label{fig:shapes}Different crystalline seeds (panels a-d) for the initialization of crystal-fluid equilibrium using $\eta=0.48$, $N=10000$ and $\eta_p^r=0.2$. The rest of the box is filled with particles from a fluid state, not shown for clarity. After $5\cdot 10^{10}$ Monte Carlo cycles a nucleus shape is obtained (panels e-h) which is very similar in all cases. Only particles belonging to the crystal (red) and the adjacent interfacial region (green) are shown. Panels (i-l) show histograms of the pressure in the surrounding fluid, and panels (m-p) show histograms of the fluid packing fraction. Within stistical accuracy these distributions are identical.}
\end{figure*}

While in the case of vapor-liquid phase coexistence in a finite simulation box in the $NVT$ ensemble one can sample the free energy of the system (and the chemical potential as function of density, see Fig. \ref{fig:chvsrho}, via a suitable numerical derivative) by umbrella sampling methods \cite{43,44,45}, this is not yet feasible for fluid-crystal phase coexistence, at least not for the desired large simulation box volumes and particle numbers. Thus we have opted for an approach where a crystal nucleus of roughly the desired size is put into a volume that is suitably cut out from a configuration of a uniform fluid phase, choosing the density of that phase (as well as the packing fraction of the crystal) roughly in the region where we expect that the system will find its equilibrium.\\
Of course, even if the chosen packing fraction of the crystal (close to $\eta_m$) and of the surrounding fluid (somewhat larger than $\eta_f$) are in the right range, so that the system can settle down to an equilibrium of the type of Eq. (13), one must make sure that this equilibrium does not depend on irrelevant details of the inital state preparation. To test this issue, we have prepared different crystalline seeds for the same value of $\eta$ for a system of $N=10000$ particles (Fig. \ref{fig:shapes}).
It is seen that the equilibrium that is established indeed is uniquely defined by the chosen parameters $N$, $V$, and $\eta_p^r$, and independent of the particular choice of initialization.\\ For the precise ``measurement'' of the density $\rho_l$ and the pressure $p_l$ of the fluid surrounding the nucleus it is important to exclude the volume region taken by the nucleus, as well as the adjacent interfacial region, from the sampling. For this purpose, we have to identify which particles belong to the nucleus, and which are definitely in the fluid region, no longer affected by the proximity of the nucleus. Following standard practice, we use the Steinhardt bond orientational parameters $q_{lm}(i)$, defined in terms of the spherical harmonics as \cite{75,76}
\begin{equation}
q_{lm}(i)=\frac{1}{N(i)}\sum_{j=1}^{N(i)}Y_{lm}\left(\theta(\vec{r}_{ij}),\phi(\vec{r}_{ij})\right)~,
\end{equation} 
where $\vec{r}_{ij}$ is the distance between the considered particle $(i)$ to one $(j)$ of its $N(i)$ neighbors within a cutoff radius which defines what the ``nearest neighbors''  are. The polar angles $\theta(\vec{r}_{ij})$ and $\phi(\vec{r}_{ij})$ of this distance vector are chosen with respect to a fixed reference frame (which is arbitrary). The cutoff radius is determined from the first minimum of the radial distribution function $g(r)$ of the bulk fcc crystal. Following \cite{76} $q_{lm}(i)$ is averaged over a particle and all its nearest neighbors,
\begin{equation}
\overline{q_{lm}}(i)=\frac{1}{N(i)+1}\sum_{k_i=0}^{N(i)}q_{lm}(k_i)~,
\end{equation} 
with $k_i=0$ meaning that one takes $q_{lm}(i)$ from Eq. (22). Averaging over $m$ we then obtain the root mean square order parameter $\bar{q_l}(i)$
\begin{equation}
\bar{q_l}(i)=\left(\frac{4\pi}{(2l+1)}\sum_{m=-l}^{+l}|\overline{q_{lm}}(i)|^2\right)^{1/2}~,
\end{equation}
relevant choices being $l=4$ and $l=6$, allowing the distinction of fcc, hcp, bcc crystals from fluids (in our study, hcp and bcc order did not play any role). From sampling the histograms $P(\bar{q_4}$, $\bar{q_6})$ both in the pure fluid and crystal phases, as well as for systems where a crystal coexists with surrounding fluid, one can easily identify suitable criteria \cite{64,73,77} to define regions in the $(\bar{q_4}$, $\bar{q_6})$-plane such that entries in these regions can be uniquely attributed to fluid-like or part of the fcc crystal, respectively.
\begin{figure}
\centering
\subfigure[]{\includegraphics[width=0.47\textwidth]{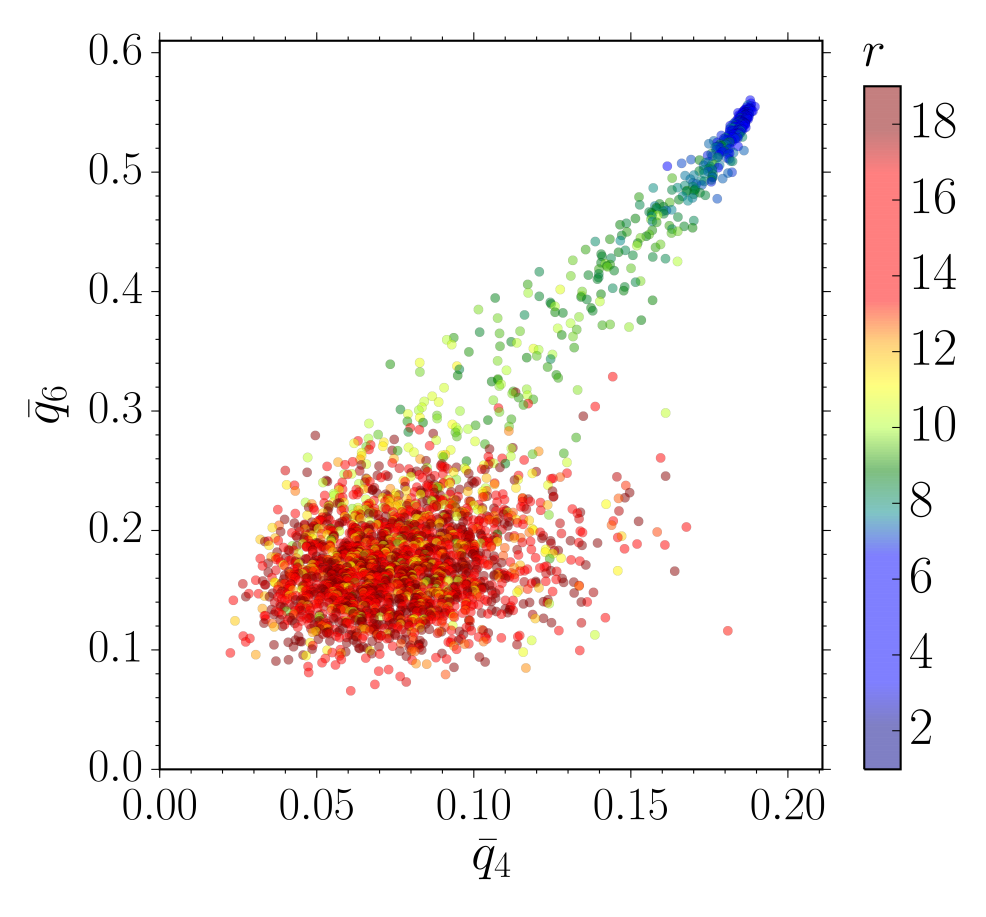}}
\subfigure[]{\includegraphics[width=0.42\textwidth]{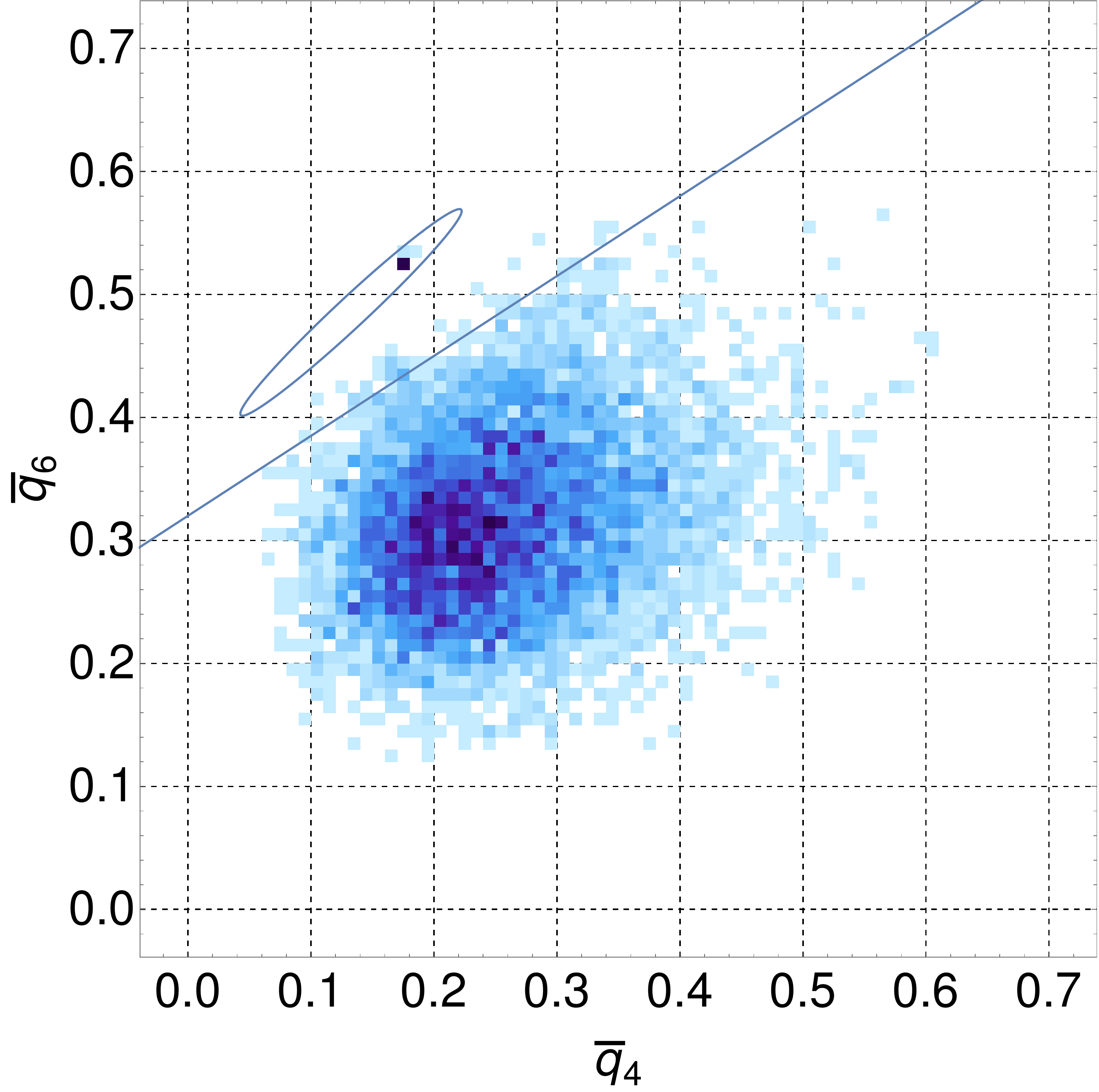}}
\caption{\label{fig:q4q6}a) Probability distribution $P(\bar{q_4},\bar{q_6})$ with a crystal nucleus at the coordinate origin shown for a system with $\eta_p^r=0.1$, $N=13000$ and $\eta_c=0.55$. Each particle in a system yields a point in this plot. The distance $r$ of each particle from the origin is shown via its color (see the scale on the rhs of the figure). Note that the broad region with $0.10\leq\bar{q_6}\leq 0.26$, $0.05\leq\bar{q_4}\leq0.20$ represents the disordered fluid, while the peak near $\bar{q_6}>0.5$, $\bar{q_4}\approx0.18$ represents the crystal. Points falling in between these regions clearly belong to the interfacial region.\\
b) Probability distribution $P(\bar{q_4},\bar{q_6})$ for the pure crystal and liquid phases at $\eta_p^r=0.28$. Only particles $i$ with $N(i)>2$ are included. Points falling below the straight line in the $\bar{q_4}-\bar{q_6}$ plane are defined as liquid, points falling into the ellipsis are defined as solid.}
\end{figure}
Of course, one can also compute approximately the crystal volume from a radial density distribution of the nucleus around its center of mass \cite{64}; while $\bar{V_n}$ defined in this way is roughly compatible with the crystal volume $\bar{V_n}$ extracted from the lever rule, Eq. (13), we emphasize that the latter method does have the clear advantage that it does not assume anything about the shape of the crystal nucleus, and thus the results that will follow are based on Eq. (13) exclusively. The analysis in terms of Eqs. (24) is only used for identifying the region in which both $\eta_l$ and $p_l$ for the fluid coexisting with the nucleus are ``measured". The result of these ``measurements'' are now shown in Fig. \ref{fig:pvseta}.
\begin{figure*}
\centering
\subfigure[]{\includegraphics[width=0.285\textwidth]{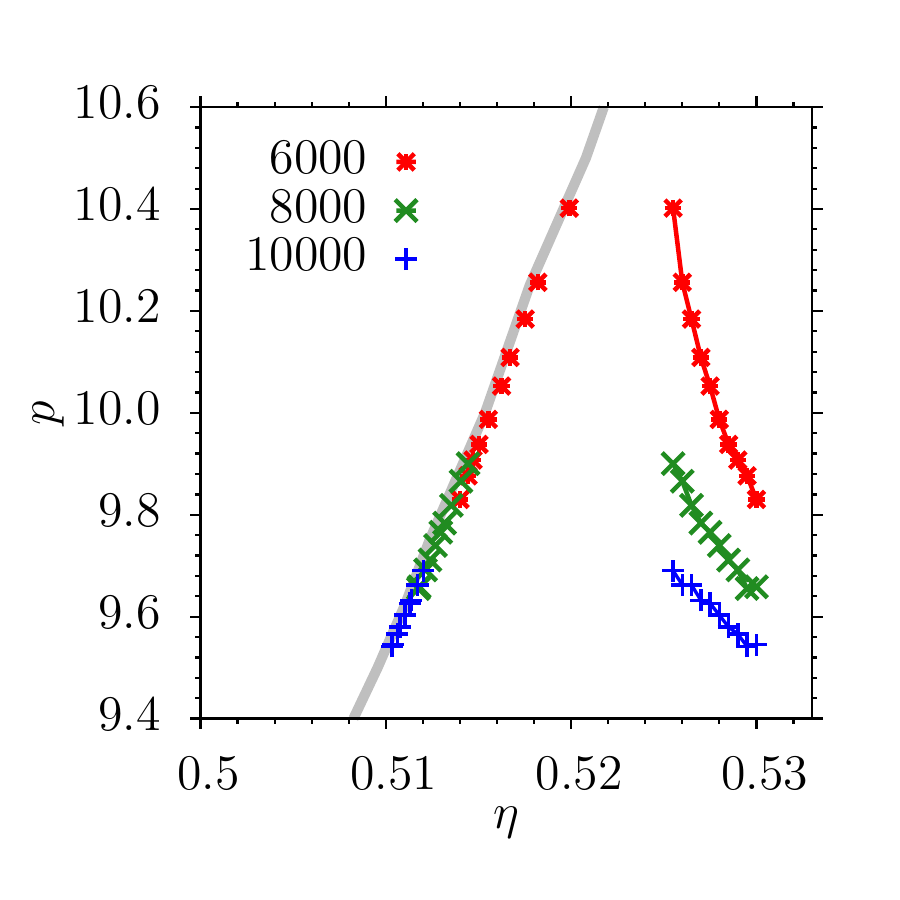}}
\subfigure[]{\includegraphics[width=0.28\textwidth]{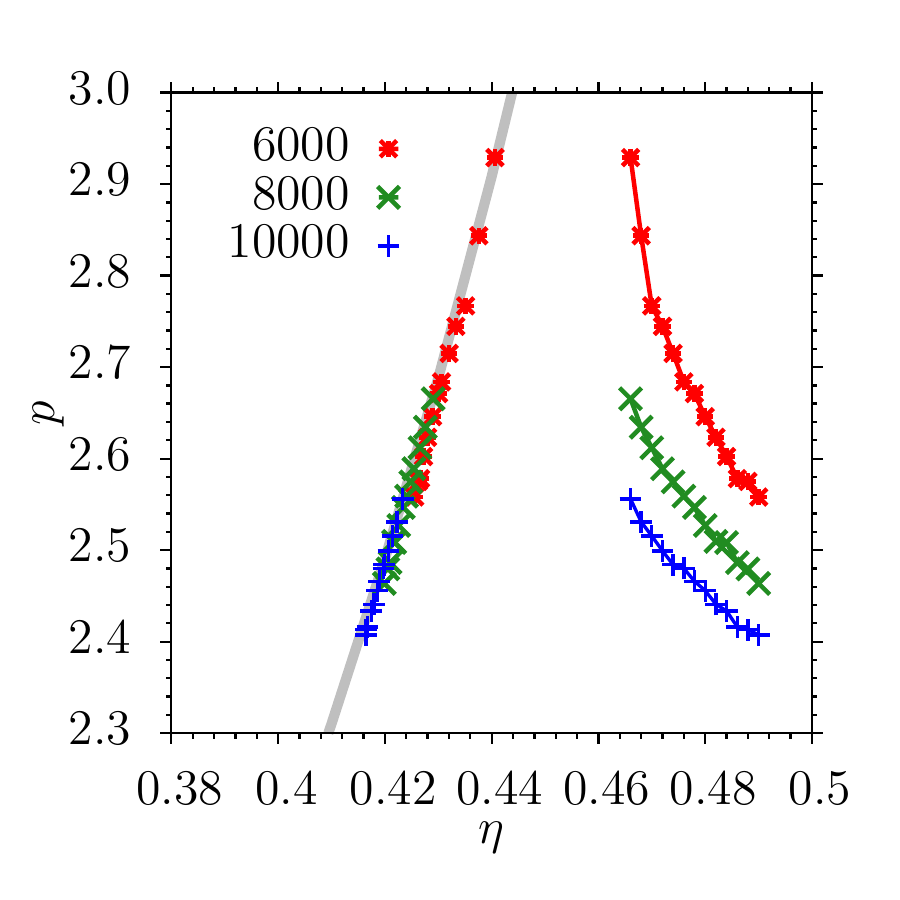}}
\subfigure[]{\includegraphics[width=0.31\textwidth]{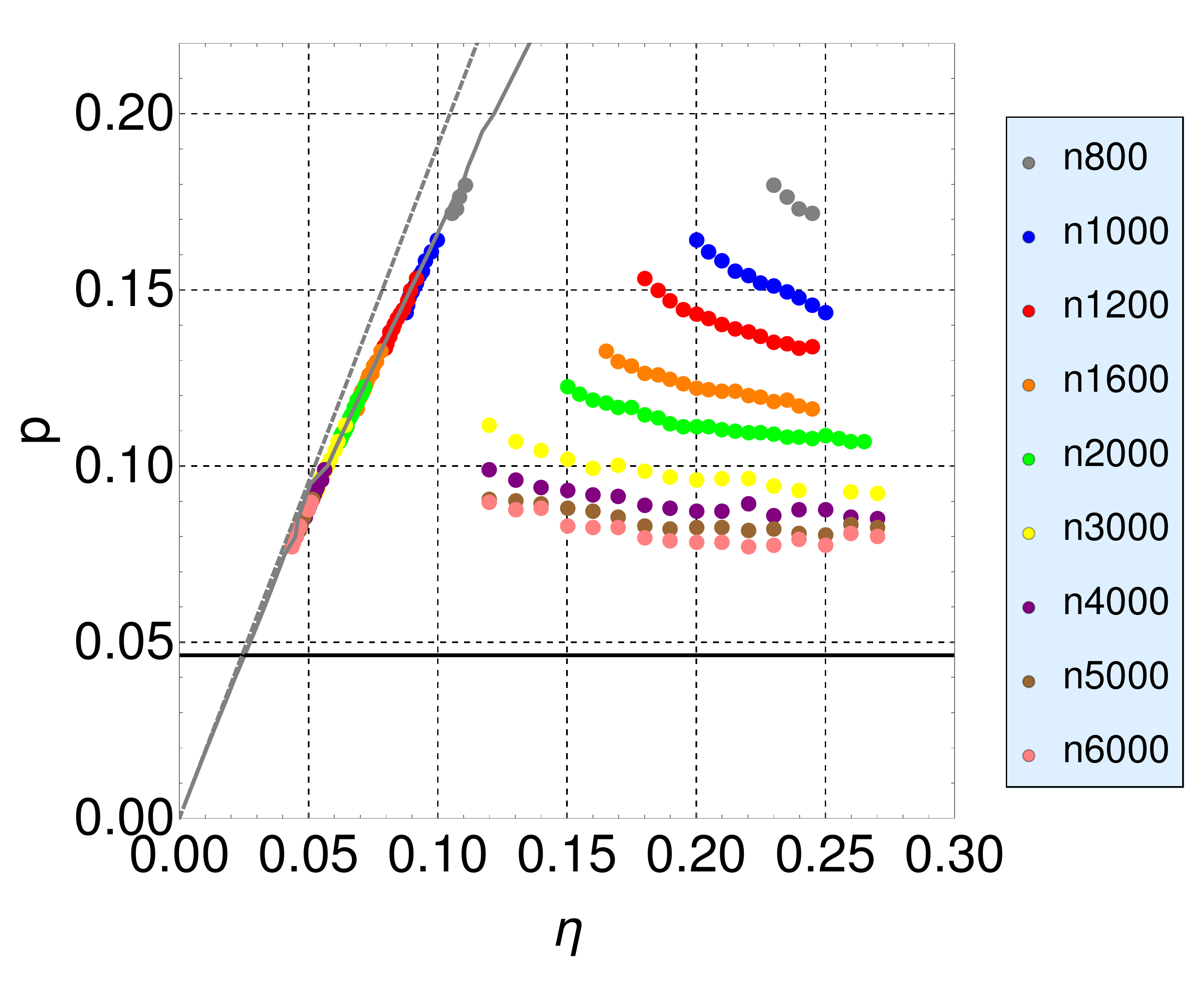}}
\caption{\label{fig:pvseta}Pressure $p_l$ of the fluid surrounding crystal nuclei plotted against the packing fraction $\eta$ in the total box (right family of curves) as well as plotted versus the actual packing fraction $\eta_l$ in this surrounding fluid (data points collapsing on the bulk equation of state, shown as a smooth curve), for $\eta_p^r=0.1$ (a), 0.2 (b) and 0.28 (c). In cases a) and b) three choices of $N$ are shown, as indicated, while in case c), 9 choices of $N$, from $N=800$ to $N=6000$, are included as indicated. Note that only a small range of $\eta$ in each case is shown. The horizontal straight line shown in case (c) indicates the estimate for $p_{coex}$, the dashed gray line indicates the ideal gas law $p(\eta)=\frac{6}{\pi}\eta$.}
\end{figure*}
One immediately recognizes that $p_l$ decreases with $\eta$, as expected from Fig. \ref{fig:chvsrho} where it was anticipated that in the region where actual coexistence of crystal nuclei with surrounding fluid occurs the chemical potential $\mu$ decreases with $\eta$, and $p_l$ is monotonously increasing with $\mu$ in equilibrium. It also is seen that the values $p_l(\eta)$ that we find are enhanced in comparison with $p_{coex}$ (Fig. \ref{fig:pvseta} c); in cases a), b) $p_{coex}$ is even beyond the scale), but this enhancement systematically decreases with increasing $N$, and our analysis implies that these enhancements must converge to zero as $N\rightarrow\infty$, and then $p_l=p_{coex}$ irrespective of $\eta$.\\
It is very instructive to study the variation of the fluid pressure $p_l$ observed in our simulations with the density (or packing fraction, respectively) observed in this fluid. These do yield the left ascending curves in the plots, and it clearly is gratifying that these data no longer depend on the choice of $N$, and simply coincide with the equation of state that has been determined from $NpT$ simulations of homogeneous fluid phases, see Sec. \ref{sec:softEffAO}. This fact implies that our choices of $N$ are large enough that finite size effects that are related to fluctuations of the bulk phases are already negligibly small. Each choice of $N$ yields data points only for a limited range of $\eta$: if $\eta$ becomes too small, the data become unreliable due to the proximity of the droplet evaporation/condensation transition; if $\eta$ becomes too large, the data become unreliable due to the proximity of the transition where the droplet shape changes from compact to cylindrical. Of course, it is an implicit assumption in Eq. (13) that the effect of such fluctuations still is negligible: if e.g. the system would exist part of the observed simulation time in a state when the droplet either has evaporated or has made a transition to the cylindrical shape or even the slab configuration, the density and pressure of the fluid would significantly change in this part of the simulation time, leading to significant deviations in the curves on the left hand side of Fig. \ref{fig:pvseta}. In fact, setting up systems for each $N$ far outside the range shown in Fig. \ref{fig:pvseta} one can verify that compact nuclei are unstable \cite{77}, and either homogeneous fluid (at small $\eta$) and cylindrical nuclei or slab configurations (at large $\eta$) actually result \cite{77}.\\
It is also evident from Fig. \ref{fig:pvseta} that the absolute values of the slopes of the decreasing part clearly decrease with increasing $N$. This property, of course, must result as the thermodynamic limit is approached: for $N\rightarrow\infty$, $p_l(\eta)=p_{coex}$ irrespective of $\eta$, and thus the slope then is zero.\\
So the qualitative features of Fig. \ref{fig:pvseta} are straightforward to understand from general principles. Using the fact that in Eq. (13) both $\rho$ and $V$ are fixed when $\eta$ and $N$ are chosen, $\rho_l=(6/\pi)\eta_l$ $\{$Eq. (17)$\}$ and the crystal density $\rho_c$ follows when we use the fact that the chemical potential $\mu_c(p_c)$ of the crystal nucleus is equal to the chemical potential of the surrounding fluid (at density $\rho_l$, pressure $p_l$, which we have separately determined, see Sec. \ref{sec:softEffAO} above). For this purpose, in general one uses Eq. (18), once $p_{coex}$ and $\mu_{coex}$ have been determined, to compute $p_c$ and extract $\eta_c(p_c)$ from the bulk equation of state (Sec. \ref{sec:softEffAO}). Note that in practice we have found that it suffices to linearize the equation of state around coexistence conditions (e.g. Eq. (19), see also Fig. \ref{fig:widom} b)). But in the case of $\eta_p^r=0.28$ it is important to take into account the nonlinear variation of $\mu$ vs $p$; note also that in Fig. \ref{fig:pvseta} c) the coexistence pressure $p_{coex}$ clearly falls in the region where the ideal gas law $p=(6/\pi)\eta$ holds, while most of the actual data for $p_l(\eta_l)$ already deviate from it. So we conclude that Eq. (13) contains a single unknown parameter, namely the nucleus volume which hence is straightforwardly extracted. We have already anticipated that knowledge of $V_n^*$ provides an alternative method, Eq. (21), based on assuming the validity of Eq. (5), and indeed Fig. \ref{fig:chvsvol} has demonstrated the selfconsistency of this procedure, although slight systematic discrepancies between the different methods to precisely locate bulk coexistence conditions remain.\\
Since we have mentioned in Sec. \ref{sec:intro}, that Eq. (2) is strictly correct only in the limit $V_n\rightarrow\infty$, and so one should discuss the effect of possible corrections that are expected to occur in Eq. (2) on our analysis. One correction that has been often identified for other systems (Ising model \cite{15,78,79}, vapor-liquid transition \cite{32,78} etc.) is an additive constant as a correction term in Eq. (2); alternatively, this can be interpreted as a curvature correction to $\bar{\gamma}$ inversely proportional to the surface area. However, it trivially follows that such a constant correction to Eqs. (2, 3) does not affect the validity of Eq. (5). 
\begin{figure*}
\centering
\subfigure[]{\includegraphics[width=0.32\textwidth]{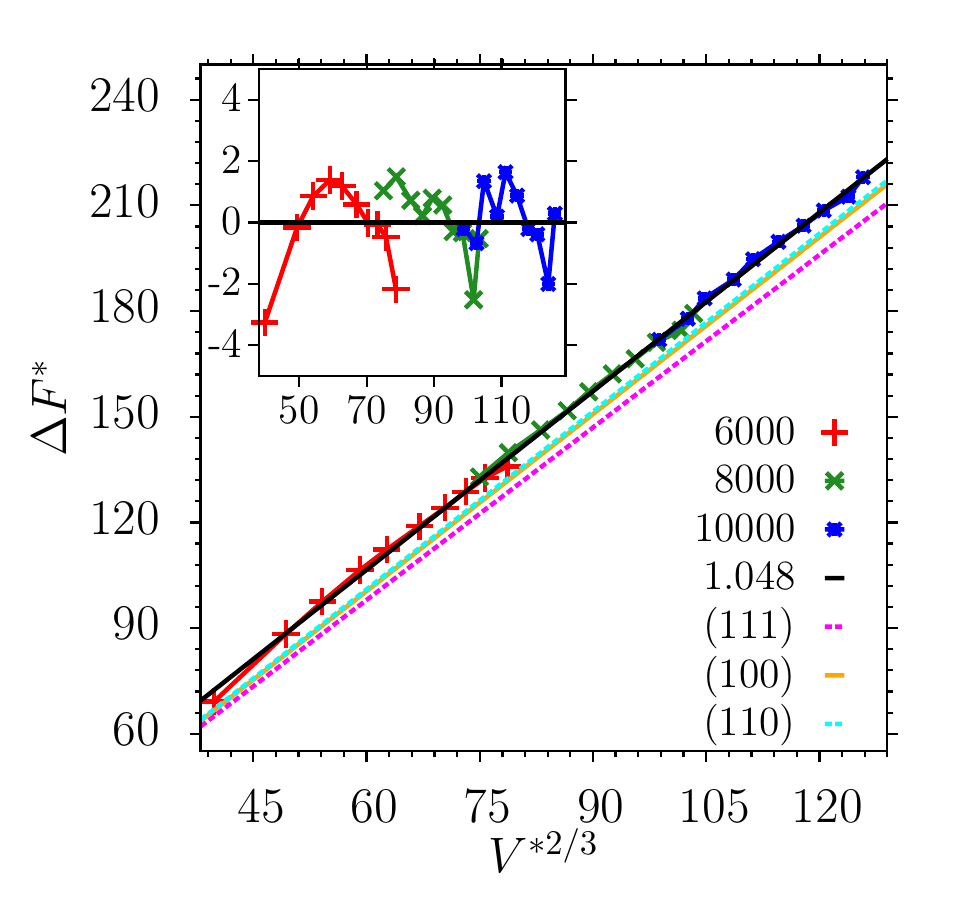}}
\subfigure[]{\includegraphics[width=0.32\textwidth]{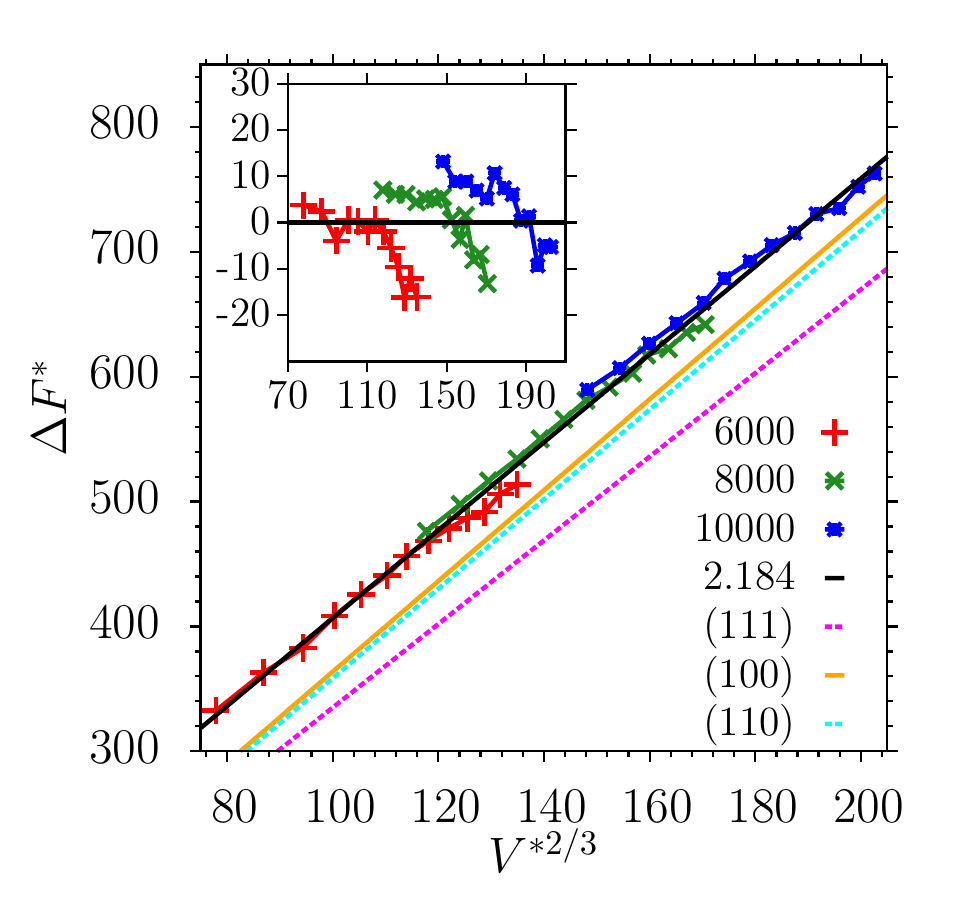}}
\subfigure[]{\includegraphics[width=0.295\textwidth]{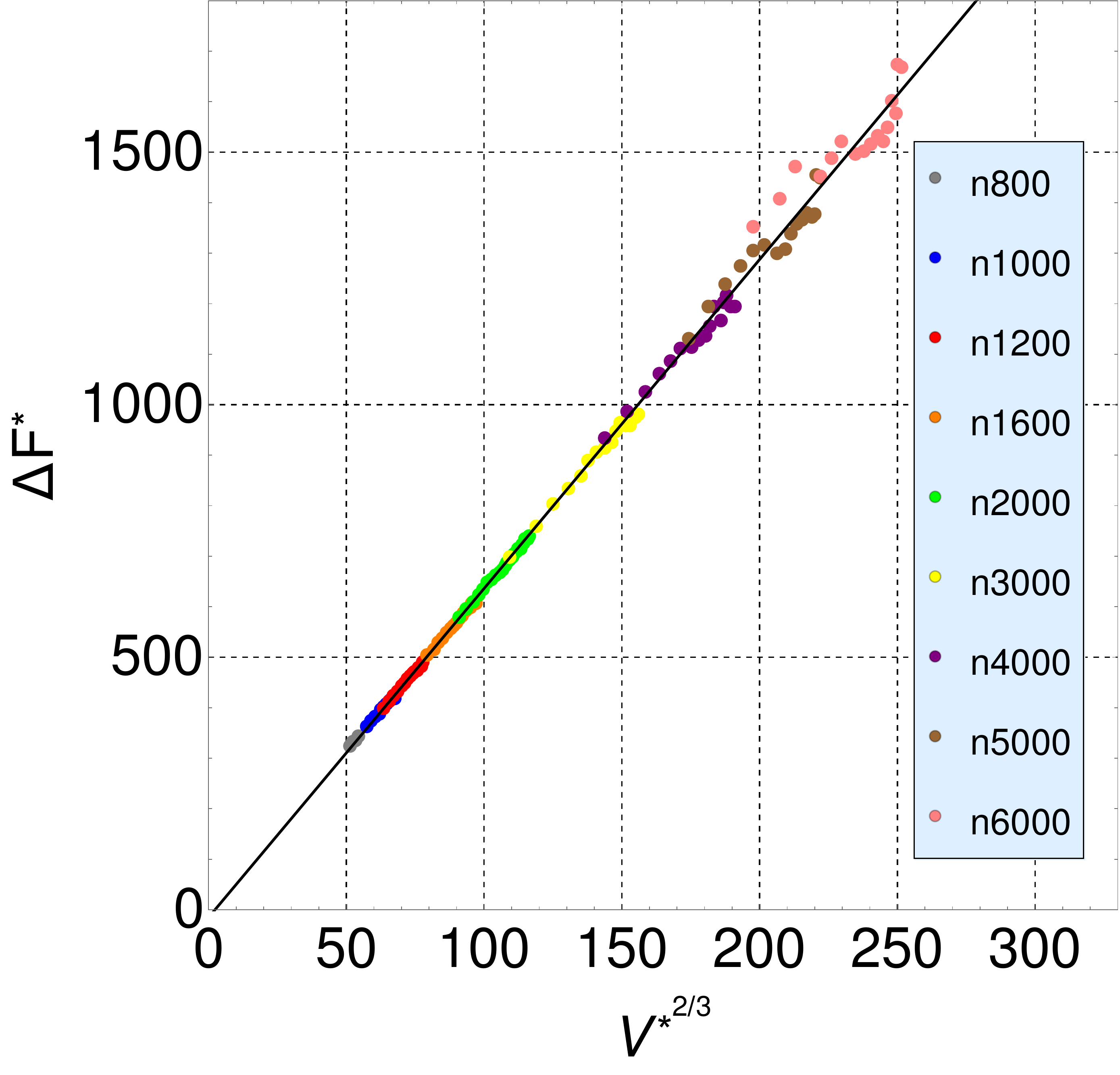}}
\caption{\label{fig:barrier}Nucleation barriers $\Delta F^*$ for the softEffAO model plotted vs. $V^{*^{2/3}}$, for $\eta_p^r=0.1$ (a), 0.2 (b) and 0.28 (c). In cases (a,b), particle numbers $N=6000$, 8000 and 10000 are used, while in case (c) particle numbers from $N=800$ to $N=6000$ are used, as indicated. Straight line fits $\Delta F^*=\left(\frac{4\pi}{3}\right)^{1/3}\gamma_{\text{eff}}V_n^{*^{2/3}}+const.$ are included, yielding $\gamma_{\text{eff}}\approx1.05$ (a), 2.18 (b) and 4.02 (c). In cases (a) and (b) also the curves $\left(\frac{4\pi}{3}\right)^{1/3}\gamma_{111}V_n^{*^{2/3}}$, $\left(\frac{4\pi}{3}\right)^{1/3}\gamma_{110}V_n^{*^{2/3}}$ and $\left(\frac{4\pi}{3}\right)^{1/3}\gamma_{100}V_n^{*^{2/3}}$ are included, where the surface tensions of the close-packed (111), (110) and (100) interfaces were independently estimated \cite{64,73}. Insets in (a) and (b) show the difference between the straight line fits and the data.}
\end{figure*}
This is no longer true for other corrections, such as fluctuation corrections leading to a $ln(V_n)$ term in Eqs. (2, 3). One then expects that for large $V_n$ Eq. (3) is modified by a correction term of relative order $(V_n^*)^{-2/3}$ and similarly in Eq. (21) a correction of order $(V_n^*)^{-1}$ appears. Unfortunately, our data are not accurate enough to identify such a term. A Tolman-type \cite{28} correction would be asymptotically larger, but since the prefactor of such corrections typically is very small \cite{28,29,30,31,32,33}, it cannot be identified either.\\
Disregarding all these corrections, we hence can use our results for $V_n$ and the associated liquid and crystal pressures $p_l$, $p_c$ to test Eq. (6). This is done in Fig. \ref{fig:barrier}.
It is seen that the data are well compatible with the relation $\Delta F^*=(4\pi/3)^{1/3}\gamma_{\text{eff}}V_n^{*^{2/3}}+const.$, as it should be for $V_n^*\rightarrow\infty$. We note that for $\eta_p^r=0.1$ the effective interfacial tension $\gamma_{\text{eff}}$ turns out to be only weakly enhanced ($\gamma_{\text{eff}}\approx 1.082$) in comparison with the actual interfacial tension $\gamma_{111}=1.013$ for the close packed (111) surface; also the surface tensions $\gamma_{110}=1.044$, $\gamma_{100}=1.039$ are not very different, and hence it is clear that the deviations from spherical nucleus shape can only be very minor, since $\gamma_{\text{eff}}=(A_W/A_{iso})\bar{\gamma}$ and $\bar{\gamma}\geq\gamma_{111}$, $A_W/A_{iso}>1$. However, for $\eta_p^r=0.2$ the anisotropy effects already are larger, $\gamma_{111}=2.078$, $\gamma_{110}=2.224$ and $\gamma_{100}=2.256$, while $\bar{\gamma}=2.406$. The surface tensions $\gamma_{111}$, $\gamma_{110}$, and $\gamma_{100}$ have been obtained elsewhere \cite{73,80,81} with the ``ensemble switch method", a thermodynamic integration method constructing a path in phase space from two $L\times L\times L_z/2$ systems with PBC throughout without interfaces, one system being fluid and the other a bulk crystal, to a system of size $L\times L\times L_z$ where the two previously separate systems were joined together, such that two interfaces are formed. Clearly, we expect that when $\eta_p^r$ increases and the difference $\eta_m-\eta_f$ also increases strongly (Fig. \ref{fig:pd}) that also the differences between $\gamma_{111}$, $\gamma_{110}$, and $\gamma_{100}$ will increase. In the limit $\eta_p^r\rightarrow\infty$ (corresponding to the limit $T\rightarrow 0$ for a molecular system), the interface tension becomes equal to the excess of the potential energy due to the interface, since then entropic contribution to the interfacial tension becomes negligible. As discussed in Sec. \ref{sec:softEffAO}, the attractive potential is strictly short range, and in the fcc crystal it acts only between nearest neighbors. For a (111) plane of the fcc crystal structure, there are 6 nearest neighbors in the plane, and 3 nearest neighbors per particle in the adjacent planes. Hence for a (111) surface, 3 nearest neighbor bonds are missing. In the pairwise energy, each bond per particle enters with a factor $1/2$, and hence the surface excess energy per particle in the (111) surface energy is $-\frac{3}{2}U(r_{min})$, in the ground state, where $r_{min}$ is the distance where the potential $U(r)=U_{\text{eff}}(r)+U_{\text{rep}}(r)$ $\{$Eqs. (14), (15)$\}$ has its minimum. Converting from the surface excess energy per particle to the surface excess energy per unit area, one concludes
\begin{equation}
U_{\text{exc}}=-\frac{\sqrt{3}}{r_{min}^2}U(r_{min})
\end{equation}
for the potential energy part of the surface tension (which is an upper bound for the surface tension, since thermal fluctuations will reduce $U_{\text{exc}}$ in comparison with Eq. (25), and a further reduction occurs due to entropic contributions to $\gamma$).
\begin{figure}
\centering
\includegraphics[width=0.35\textwidth]{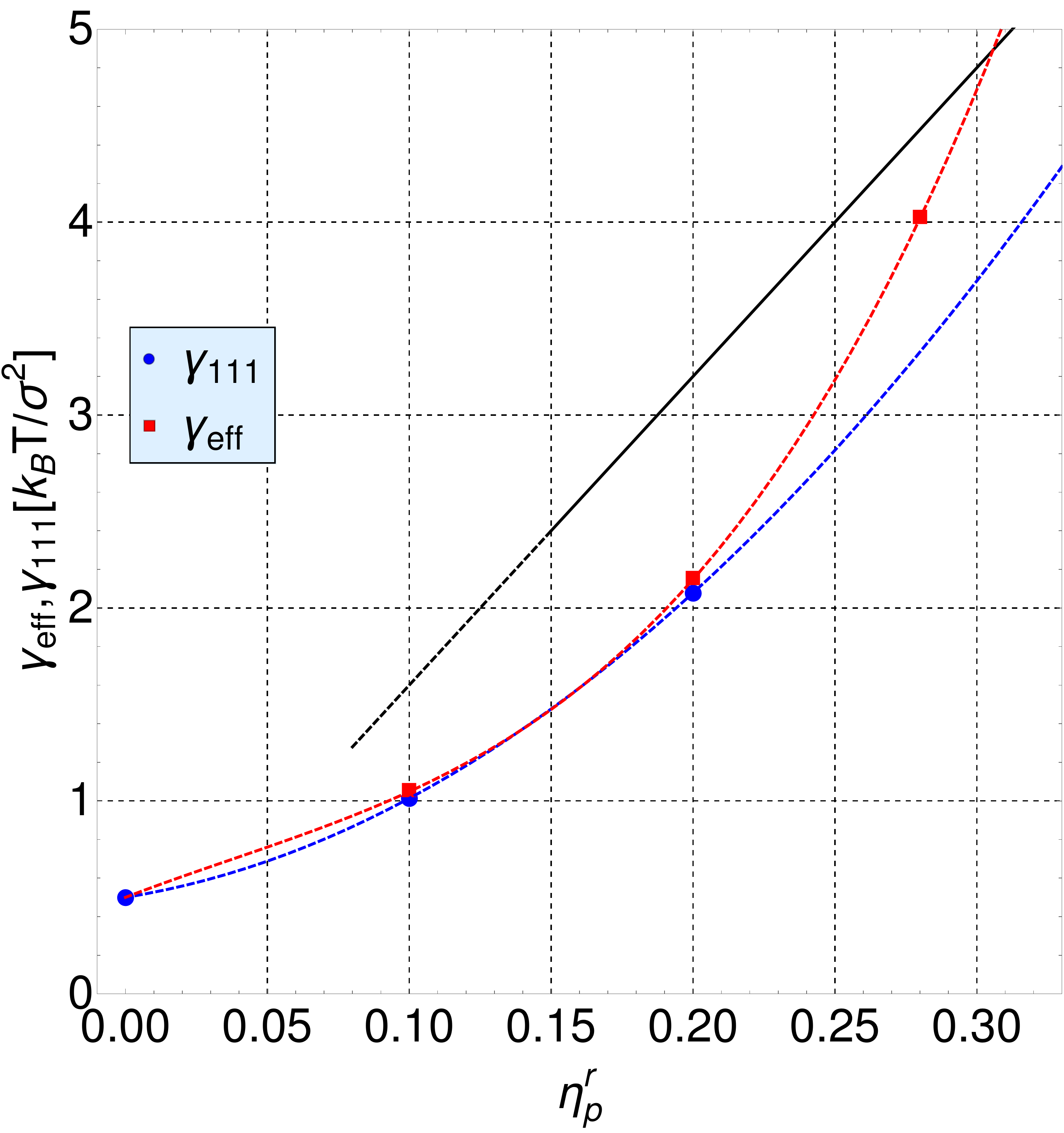}
\caption{\label{fig:estimate}Effective interfacial tension $\gamma_{\text{eff}}$ (red) with polynomial (dashed red curve, guide to the eyes only), interface tension $\gamma_{111}$ of a close-packed (111) surface (blue) with polynomial (dashed blue curve, guide to the eyes only), and its upper bound $U_{\text{exc}}$ $\{$Eq. (25), black line$\}$ plotted vs $\eta_p^r$.}
\end{figure}
But one sees that for $\eta_p^r=0.28$ the upper bound for $\gamma_{111}$ is already rather close to $\gamma_{\text{eff}}$, suggesting that for still larger $\eta_p^r$ this upper bound for $\gamma_{111}$ can already be used to obtain order of magnitude estimates for the actual surface tension. Of course, one must keep in mind that $\gamma_{\text{eff}}$ deviates from $U_{\text{exc}}$ due to two competing effects: $\gamma_{\text{eff}}$ is higher than $\gamma_{111}$ due to anisotropy effects; but there is a reduction of the surface tension (in comparison to $U_{\text{exc}}$) due to entropy \cite{57a}.\\
Finally, Fig. \ref{fig:snap} shows a typical snapshot of a crystal nucleus for $\eta_p^r=0.28$.
\begin{figure}
\centering
\subfigure[]{\includegraphics[width=0.47\textwidth]{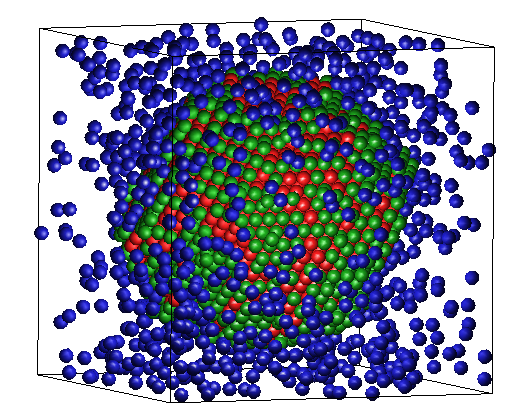}}
\subfigure[]{\includegraphics[width=0.45\textwidth]{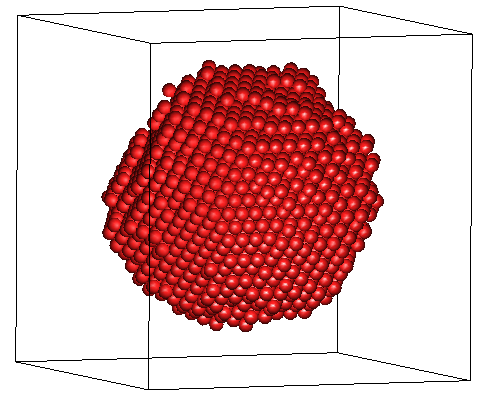}}
\caption{\label{fig:snap}Typical snapshot picture of a crystal nucleus at $\eta_p^r=0.28$, for $N=5000$ particles and $\eta=0.21$. Case a) shows all particles in the system (fluid particles are indicated in blue, interface particles are indicated in green, crystal particles are indicated in red), while case b) shows the nucleus only.}
\end{figure}
Although it clearly does not a have a perfectly regular facetted shape, it is not a rough spherical shape either.
\section{Conclusions\label{sec:conc}} 
In this paper a method has been described by which nucleation barriers for homogeneous nucleation of crystals from the fluid phase can be computed, emphasizing the limting case where the barrier is due to the competition between the difference in bulk thermodynamic potential between the fluid and solid phases with the surface free energy of a crystal nucleus which has the Wulff equilibrium shape when the crystal nucleus is large enough. It has been shown that the problem of finding this shape can be bypassed, analyzing the equilibrium properties (pressure, density, chemical potential) of the fluid coexisting with the crystal nucleus in a finite volume already contains all the information that is required to compute the nucleation barrier. This concept is exemplified for a simple model of colloidal particles with a short range effective attraction (caused physically by the depletion effect due to polymers, for instance). Choosing the size ratio of polymers and colloids $q=0.15$, the effective colloid-colloid attraction is strictly short-range and pairwise. In this model the fluid density where crystallization starts (or associate packing fraction $\eta_f$, respectively) is close to its counterpart in the crystal where melting starts ($\eta_m$) if the attraction is small, but the difference $\eta_m-\eta_f$ can be continously increased up to the limit where $\eta_f\rightarrow 0$ and $\eta_m$ approaches the packing fraction of close-packed spheres. The chosen softEffAO model can be studied by Monte Carlo simulation both in the $NVT$ and in the $NpT$ ensemble very efficiently; but the fluid-solid transition typically suffers from hysteresis, and very accurate location of coexistence conditions (pressure $p_{coex}$, chemical potential $\mu_{coex}$, etc.) is still one of the main factors limiting the accuracy of the present work. However, we feel that the situation is far more favorable than in experimental studies of colloidal systems, where neither the interaction potential nor the properties of the colloids themselves are known with sufficient accuracy to perform an experimental counterpart of our study (e.g., polydispersity of the colloids matters, ions in the solution adsorbing on the particle surface change the interactions slightly, etc.). Thus the accuracy with which experimental phase diagrams are measured does not match data such as shown in Figs. 4, 5; it is hoped that experimentalists will take up this challenge.\\
In our work, we have found that the conventional theory of homogeneous nucleation works nicely, the nucleation barriers $\Delta F^*(V_n^*)$ scale with the nucleus volume $V_n^*$ as $\Delta F_n^*=(\frac{4\pi}{3})^{1/3}\gamma_{\text{eff}}V_n^{*^{2/3}}+const.$, $\gamma_{\text{eff}}$ being slightly but systematically enhanced in comparison with the surface tension $\gamma_{111}$ of close-packed (111) surfaces in the fcc crystal lattice. Experimental work often uses a similar expression, but $\gamma_{\text{eff}}$ is taken as a fit parameter that depends distinctly on the fluid density. We do not need to make such unjustified ad hoc assumptions here.\\
For the range of the free energy barriers for which we confirm the validity of the classical theory of homogeneous nucleation in general, namely $10^2k_BT\leq\Delta F^*\leq 10^3 k_BT$, homogeneous nucleation is not directly observable, of course. However, we note that for many instances of heterogeneous nucleation the corresponding barriers are much smaller, and can only be predicted if the barriers for homogeneous nucleation are known. E.g., when one considers heterogeneous nucleation on planar substrates for conditions of partial wetting, the corresponding barriers are $\Delta F_{het}^*=\Delta F_{homo}^*~f_{VT}(\theta)$, where the reduction factor $f_{VT}(\theta)$ predicted by Vollmer and Turnbull is of order $f_{VT}\approx10^{-1}$ if the contact angle is small. In any case, the knowledge of what the classical theory would predict also is a useful point of reference under conditions where the classical theory does not hold.\\
The present approach should be of general applicability, irrespective of the type of material that is studied. Thus, we hope to report on further applications, such as nucleation of ice crystals from supercooled water, in the future.\\\\
\begin{acknowledgements} Partial support from the Deutsche Forschungsgemeinschaft (grant VI 237/4-3) and the Graduate School of Excellence Materials Science in Mainz is acknowledged. P. Koß is a recipient of a DFG-fellowship/DFG-funded position through the Excellence Initiative by the Graduate School Materials Science in Mainz (GSC
266). The authors gratefully acknowledge the computing time granted on the supercomputer Mogon at Johannes Gutenberg University Mainz (hpc.uni-mainz.de) and Hornet at the Höchstleistungsrechenzentrum Stuttgart (HLRS). 
\end{acknowledgements}

\end{document}